\documentclass[aps,prd,twocolumn,groupedaddress, nofootinbib, longbibliography,superscriptaddress,amsmath,amssymb,amsfonts, floatfix]{revtex4-2}
\usepackage{graphicx} 
\usepackage[dvipsnames]{xcolor}
\usepackage{hyperref}

\definecolor{mydarkblue}{RGB}{46, 48, 146}
\hypersetup{colorlinks=true,allcolors=mydarkblue}
\usepackage{multirow}
\usepackage{adjustbox}
\usepackage[utf8]{inputenc} 
\usepackage[T1]{fontenc}    
\usepackage{microtype}      
\usepackage{tensor}
\usepackage{orcidlink}
\usepackage[all]{hypcap}
\usepackage{subfigure}
\usepackage[capitalise]{cleveref} 
\usepackage{url}
\usepackage{orcidlink}
\usepackage{booktabs}
\usepackage{multirow}
\usepackage{aas_macros}
\usepackage{cancel}
\usepackage[normalem]{ulem}

\begin{document}

\title{Robustness of Neural Networks for CMB Polarization Foreground Removal}

\author{\mbox{Luca Gomez Bachar}\orcidlink{https://orcid.org/0009-0002-2942-2162}}
\email{luca.gomez@iteda.gob.ar}
\affiliation{Instituto de Tecnologías en Detección y Astropartículas (CNEA, CONICET, UNSAM),
Centro Atómico Constituyentes, San Martín, Buenos Aires CP B1650KNA, Argentina}
\author{\mbox{Cora Dvorkin}\orcidlink{}}
\affiliation{Department of Physics, Harvard University, Cambridge, MA 02138, USA}
\author{\mbox{Alberto Daniel Supanitsky}\orcidlink{}}
\affiliation{Instituto de Tecnologías en Detección y Astropartículas (CNEA, CONICET, UNSAM),
Centro Atómico Constituyentes, San Martín, Buenos Aires CP B1650KNA, Argentina}

\date{\today}

\begin{abstract}
The detection of Cosmic Microwave Background primordial $B$-mode polarization would constitute a ``smoking gun" signal of primordial gravitational waves. However, this measurement requires accurate removal of polarized Galactic foregrounds to avoid systematic biases when estimating the tensor-to-scalar ratio. Methods based on Machine Learning (ML) techniques, such as Convolutional Neural Networks (CNNs), have recently been proposed as alternative foreground cleaning techniques, but their applicability to real data relies on their ability to generalize beyond the models assumed during training. In this work, we focus on a variety of foreground models (FMs) used for training and conduct a systematic study of the generalization properties of a CNN-based method. We train two CNN architectures on simulations generated from different Galactic FMs, and test their performance on models not used during the training. By characterizing the pixel-to-pixel statistical properties of the FMs using variance, skewness, and Shannon entropy, we define a marginal statistical complexity hierarchy among them. The morphology diversity of the models is also incorporated in the hierarchy established. We show that training on the more complex FMs reduces bias and improves precision when testing on unseen template-based FMs, whereas training on the simplest model could introduce systematic errors, reaching a residual equivalent to $r>0.1$ on both models used only for testing. We further test this picture by constructing controlled modifications of the simplest FM with progressively broader marginal distributions. While their marginal statistics grow beyond those of the most morphologically diverse model, their generalization improves only up to a point and none of them matches the robustness across test models of the more complex model, showing that marginal variability alone is not sufficient for generalization. These results evidence that a lack of generalization is a relevant source of systematic uncertainty, and emphasize the importance of understanding the impact of the models assumed during training in ML-based methods before applying them to real data.
\end{abstract}

\maketitle

\section{Introduction}

The standard cosmological model ($\Lambda$CDM) provides a successful description of the evolution of the universe. Nevertheless, several fundamental questions remain open, some of them regarding the physics of the very early Universe. Inflationary scenarios were originally introduced to explain conceptual challenges such as the horizon and flatness problems and, more generally, to provide a mechanism for generating primordial perturbations \cite{guth1981inflationary, LINDE1982389}. A generic prediction of inflation is the existence of a stochastic background of primordial gravitational waves (PGW), whose detection would provide the first observational window into physics at energy scales far beyond the reach of terrestrial experiments.

Tensor perturbations of the metric give rise to a non-vanishing PGW power spectrum. These tensor modes leave a distinctive imprint on the polarization of the Cosmic Microwave Background (CMB), sourcing both $E$- and $B$-mode polarization, while scalar perturbations under linear evolution generate only $E$-modes. As a consequence, a primordial $B$-mode signal constitutes a direct probe of inflationary physics and is commonly parametrized by the tensor-to-scalar ratio $r$. A detection of $r>0$ would therefore represent a smoking gun for the inflationary epoch. Currently, the best constraint is obtained from the measurements by the BICEP/Keck
Collaboration \cite{ade2021improved} reaching $r<0.032$, when combined with the Planck Release 4 (PR4) and Baryon Acoustic Oscillations \cite{Tristram_2022}.

Measuring the primordial $B$-mode signal is challenging. In addition to instrumental noise \cite{tegmark1996method} and systematic effects associated with the optical design of CMB experiments \cite{hu2003benchmark,stellati2025systematic}, weak gravitational lensing converts $E$-modes into $B$-modes \cite{lewis2006weak}, producing a secondary signal that must be mitigated through delensing techniques \cite{belkner2024cmb}. Furthermore, polarized Galactic foregrounds, which are dominated primarily by synchrotron radiation and thermal dust emission, among other polarized sources, contaminate the observed CMB maps and can mimic the primordial $B$-mode signal \cite{ade2015joint,akrami2020planck,martire2022characterization,choi2015polarized}. An inaccurate treatment of these foregrounds can lead to biased estimates of $r$, making foreground removal a dominant source of systematic uncertainty in current and future CMB polarization experiments \cite{bianchini2025cmb}.

Several foreground-cleaning techniques have been developed to address this problem. Parametric approaches rely on explicit modeling of the frequency dependence of foreground components \cite{Stompor_2009,poletti2023fgbuster,abazajian2022cmb}, while blind methods derived from the Internal Linear Combination (ILC) aim to minimize foreground contamination without strong assumptions of the Foreground Model (FM) \cite{bennett1992preliminary, bennett2003first, tegmark2003high, eriksen2004foreground,delabrouille2009full, remazeilles2011cmb, leloup2023nonparametric}. Although traditional methods have achieved considerable success, their performance could be insufficient to reach the sensitivity required for detecting primordial $B$-modes in forthcoming experiments \cite{bianchini2025cmb}, motivating the exploration of alternative approaches.

In recent years, machine learning (ML) techniques have been applied to several fields in cosmology and astrophysics \cite{Gomez_Bachar_2025,kannan2025cosmoflow,zeghal2025simulation,chen2025field, akhmetzhanova2024data,10.1093/mnras/stad3521,sun2025conditionalvariationalautoencoderscosmological}. In particular, several ML-based approaches have been proposed as tools for CMB component separation, such as Convolutional Neural Networks (CNNs) techniques \cite{casas2022cenn,wang2022recovering, yan2023recovering, yan2024cmbfscnn}, Diffusion Models (DMs) \cite{heurtel2023removing}, and other approaches which are not based on image processing, like ML-based ILC extensions \cite{mccarthy2025signal,yadav2024perceptronbasedilcmethod,duivenvoorden2025robust}. 

In this work, we study the generalization properties of a CNN-based foreground-cleaning method applied to CMB polarization, adopting as a case study a CMB-S4-like small aperture telescope (SAT) \cite{abitbol2017cmb,abazajian2016cmb}. Rather than assuming that neural networks automatically generalize across different FMs, we explicitly test this assumption by training CNNs on simulations generated from specific template-based Galactic FMs and evaluating their performance on other models not seen during the training. To this end, we introduce performance metrics to quantify both the accuracy and precision of the reconstructed CMB signal and to identify potential biases and uncertainty arising from the lack of generalization.

To interpret these results, we quantify statistical properties (variance, skewness, and Shannon entropy) of the FMs considered in this work. We show that these statistics serve as a measure of foreground complexity. In this sense, generalization emerges not merely as a machine learning concern but as a physically relevant property for controlling the systematic uncertainty in the estimation of cosmological parameters.

This paper is organized as follows. In Section~\ref{sims} we describe the simulation pipeline used to generate CMB polarization maps, instrumental noise, and Galactic foregrounds. In Section~\ref{methods} we present the foreground-cleaning methods considered in this work, including the traditional techniques and the CNN-based approach. In Section~\ref{gen} we introduce our generalization hypothesis, compute the statistical properties of the FMs, and define the performance metrics used in the analysis. Finally, in Section~\ref{res} we present and discuss our results.

\section{Simulations}\label{sims}

The goal of this paper is to analyze the impact of the simulation-based technique on the performance of the CNN-based method. To this end, we define a procedure to generate simulations to train and test the networks, including different FMs. We provide the code used here\footnote{\url{https://github.com/LucaGomez/cNNbP}}. Our simulations include three components: lensed CMB polarization maps, instrumental noise realization maps, and polarized Galactic foregrounds. To generate our simulations, we use the Python library Healpy\footnote{\url{https://github.com/healpy/healpy}} \cite{2005ApJ...622..759G, Zonca2019}, and for the simulations used to train and test the networks we adopt $N_{\rm side}=256$, leading to a $\ell_{\rm max}\approx767$, which is enough to target the recombination bump at $\ell\approx100$. 
We perform the simulations using the spin-two polarization Stokes Parameters $Q$ and $U$, which are physical quantities that can be measured directly by a telescope.   

\begin{figure}
    \centering
    \includegraphics[width=\linewidth]{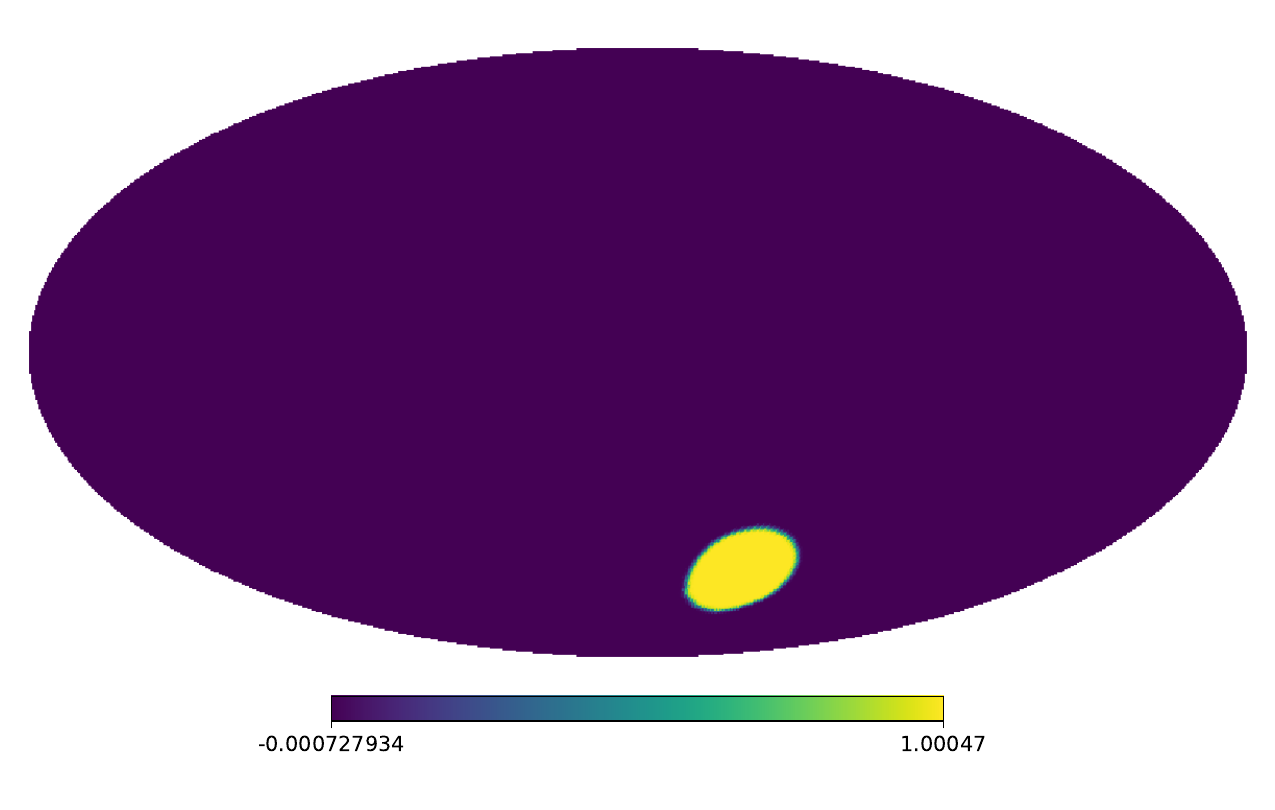}
    \caption{Apodized version of the mask applied to the full-sky simulations. The observed patch is centered at Galactic coordinates $(l, b) = (316^\circ, -56^\circ)$}. The apodization scale applied in this mask is $0.5^\circ$.
    \label{mask}
\end{figure}

As our analysis focuses on ground-based experiments, we use a partial-sky approach. To account for that, we adopt a mask centered at the Galactic coordinates $(l, b) = (316^\circ, -56^\circ)$, a region at high Galactic latitude. We use an angular aperture with a radius of $14^\circ$ centered on this point. To compute the angular spectra, we adopt an apodized version of this mask, which is shown in Fig.~\ref{mask}, with an apodization scale of $0.5^\circ$. The sky patch studied in this work corresponds to a sky fraction of $f_{\rm sky} \approx 0.015$.

For the CNN-based method, we generate simulated training, validation, and testing sets with $1000$, $300$, and $300$ polarization maps, respectively. For each simulation, we generate the lensed CMB, the instrumental noise, and the Galactic foregrounds. In the rest of this section, we detail the process for generating the components mentioned above. 

In this work, we consider only dust and synchrotron emission, since the scope of this paper is the generalization capabilities of the CNN-based method, and we begin the analysis with a simple setup that includes only these two foreground sources together with the CMB. This choice limits the scope of our results: additional sources, such as anomalous microwave emission and polarized point sources, among others, would need to be included before applying this method to real data.

\subsection{Lensed CMB polarization maps}

To generate lensed realizations of CMB polarization maps, we start by computing the primordial power spectra using CAMB\footnote{\url{https://github.com/cmbant/CAMB}} \cite{lewis2000efficient} assuming a fixed cosmology. We adopt the Planck 2018 best-fit cosmological parameters \cite{aghanim2020planck}, fixing $r=0$ for all the simulations in this work. These primordial power spectra are then used to generate lensed realizations of the $Q$ and $U$ CMB polarization maps with Lenspyx\footnote{\url{https://github.com/carronj/lenspyx}} \cite{Reinecke_2023}. Since we work in units of $\mu$K$_{\rm CMB}$, we use the same lensed CMB polarization maps for all frequency channels.  
We compute the pseudo-$C_\ell$s of the lensed CMB polarization maps using the Python library Namaster\footnote{\url{https://github.com/LSSTDESC/NaMaster/}} \cite{alonso2019unified}. To perform this computation, we apply the mask shown in Fig~\ref{mask} to the $Q$ and $U$ maps. We then use {\rm Namaster} to estimate the pseudo-$C_\ell$, including $B$-mode purification \cite{grain2009polarized}, and adopting a binning scheme of 12 multipoles per bin. The pseudo-$C_\ell$s computed following this procedure are used as the ground-truth, $C_\ell^{\rm True}$, to measure the performance of the methods in the analysis of the results. For the CNN, these maps are the CMB at 220~GHz,
convolved with the beam of that channel and include its instrumental noise realization.

\subsection{Instrumental noise}

To simulate instrumental noise, it is necessary to use a telescope model. Since our goal is to target primordial $B$-modes at $\ell \approx 100$, corresponding to the recombination bump, we adopt a SAT model. Several collaborations use SATs to measure large angular scales, targeting the recombination bump of the B-mode of polarization. Some current examples are the Simons Observatory \cite{galitzki2024simons} and BICEP \cite{yoon2006robinson,schillaci2020design}, and future ones are QUBIC \cite{hamilton2022qubic,mousset2022qubic,torchinsky2022qubic} and CMB-S4 \cite{abazajian2016cmb}. In this paper, we adopt the CMB-S4 SAT model for simplicity. 
The CMB-S4 SAT measures each frequency band with independent detectors \cite{abazajian2016cmb}, allowing us to safely assume an uncorrelated noise model across frequency bands.

We therefore generate Gaussian noise realizations for all the frequency bands considered, assuming a $1/f$-type noise power spectrum:

\begin{equation}\label{1/f}
    N_\ell^{X,\nu} = \left(\frac{\pi\,\Delta^\nu}{10800}\right)^2\left(1+\left(\frac{\ell}{\ell_{\rm knee}^\nu}\right)^{\alpha_{\rm knee}^\nu}\right),
\end{equation}
where $X=\{E,B\}$, and the parameters of the noise power spectrum can differ between frequency bands \cite{yan2023recovering,belkner2024cmb,bianchini2025cmb}. The parameters assumed for the SAT telescope are listed in Table~\ref{SATmodel} \cite{belkner2024cmb}. To generate Gaussian noise realizations, we use the Synfast function from Healpy. This approach transforms the full-sky $E$- and $B$-mode realizations generated with the spectrum from Eq.~\ref{1/f} into full-sky $Q$ and $U$ maps. After this transformation, we apply the mask to get the partial-sky noise realization. This noise model corresponds to four years of observations taken by the adopted SAT model \cite{abazajian2022cmb}.

\begin{table}[h]
\centering
\begin{tabular}{c|c|c|c|c|c}
\toprule
 & Band & Beam & \multicolumn{3}{c}{Spectra parameters}  \\
SAT & (GHz) & (arcmin) & $\Delta^\nu$ ($\mu$K arcmin) & $\ell_{\rm knee}^\nu$ & $\alpha_{\rm knee}^\nu$  \\
\midrule
 & 30  & 72.8 & 3.74 & 60 & $-2.2$  \\
 & 40  & 72.8 & 4.73 & 60 & $-2.2$  \\
 & 85  & 25.5 & 0.93 & 60 & $-2.2$  \\
 & 95  & 22.7 & 0.82 & 60 & $-2.2$  \\
 & 145 & 25.5 & 1.25 & 65 & $-3.1$  \\
 & 155 & 22.7 & 1.34 & 65 & $-3.1$  \\
 & 220 & 13.0 & 3.48 & 65 & $-3.1$  \\
 & 270 & 13.0 & 8.08 & 65 & $-3.1$  \\
\bottomrule
\end{tabular}
\caption{Small aperture telescope model assumed for the simulations.}
\label{SATmodel}
\end{table}

\subsection{Foreground models}\label{FM}

The foreground models studied in this work are based on templates derived from observations. We consider three models for training the CNNs. The first one is a model based on Gaussian fluctuations of the spectral energy distribution (SED) parameters, which we call ``GP"; the second model consists of rotating the sky before extracting the sky patch, and we call it ``RP"; and the third one is the PySM3\footnote{\url{https://github.com/galsci/PySM}} model d11s6 \cite{Zonca_2021}. The three FMs are based on templates derived from observations, which are available in PySM3 and generate full-sky realizations across all frequency bands. Our Galactic FMs include only dust and synchrotron emission, which is sufficient for the aims of this paper; additional polarized sources can be included using a similar procedure. For the FMs, we assume the same SED \cite{Planck2016}: a modified blackbody (MBB) spectrum for dust:

\begin{equation}\label{MBB}
\begin{aligned}
Q_\nu^{d}(\hat n)
&=
A_{Q,\nu_0}^{d}(\hat n)
\left(\frac{\nu}{\nu_0}\right)^{\beta_d(\hat n)}
B_\nu\!\left(T_d(\hat n)\right),
\\[6pt]
U_\nu^{d}(\hat n) 
&=
A_{U,\nu_0}^{d}(\hat n)
\left(\frac{\nu}{\nu_0}\right)^{\beta_d(\hat n)}
B_\nu\!\left(T_d(\hat n)\right),
\end{aligned}
\end{equation}
where $B_\nu\!\left(T_d(\hat n)\right)$ is the typical black body energy density at frequency $\nu$ for a temperature $T_d$ at the direction in the sky described by $\hat{n}$. $A_{Q,\nu_0}^{d}$ and $A_{U,\nu_0}^{d}$ are the amplitudes, $\beta_d$ is the spectral index, and $\nu_0=353$ GHz is the pivot frequency for polarization. For the synchrotron emission, we consider a power law dependence:

\begin{equation}\label{powerlaw}
\begin{aligned}
Q_\nu^{s}(\hat n)
&=
A_{Q,\nu_0}^{s}(\hat n)
\left(\frac{\nu}{\nu_0}\right)^{\beta_s(\hat n)},
\\[6pt]
U_\nu^{s}(\hat n) 
&=
A_{U,\nu_0}^{s}(\hat n)
\left(\frac{\nu}{\nu_0}\right)^{\beta_s(\hat n)}.
\end{aligned}
\end{equation}

In addition to the three models presented above, we use two additional models implemented in PySM3, d4s2 and d12s7, only for testing the CNNs. In the following sections, we describe the details about each FM studied in this work.

\subsubsection{Gaussian fluctuations in the SED parameters (GP)}

This model is based on the PySM3 model d1s1. For the dust contribution (d1), the parameters are based on Planck 2015 observations at 353~GHz for polarization \cite{Planck2016}. These observations are scaled to different frequencies assuming an MBB SED and using the spatially varying temperature and spectral index obtained from the Planck data using the Commander code 
\cite{Planck2016}. For the synchrotron (s1), the templates are based on WMAP 9-year 23 GHz Q/U maps \cite{bennett2013nine} and are scaled to different frequencies using a power-law assumption. To generate variations in the fixed d1s1 template, we introduce fluctuations in the parameters appearing in Eqs.~\ref{MBB} and \ref{powerlaw}, which affect the frequency scaling of the original template. These variations are introduced following the technique described in \cite{yan2023recovering, yan2024cmbfscnn}. This procedure introduces 5\% variations in $\beta_d$ and $\beta_s$, 10\% variations for the amplitudes, and 5\% variations for the dust temperature in each realization. 

We also consider wider variations of GP referred as GP-Wx, where x refers to a factor that amplifies the variations in the SED parameters. The variations considered are GP-W3 and GP-W8, being the variations for these models 15-30\% and 40-80\% respectively. We also consider GP-W8M, which corresponds to variations of 40-80\% and samples a different value for each pixel, while all the previous models samples a unique value for the whole map of each parameter, except the dust temperature, which varies from pixel to pixel.

\subsubsection{Rotating the sky before extracting the patch (RP)}

This foreground model is based on the d1s2 template. The dust contribution is the same as in the GP model, and the synchrotron spectral index steepens off the Galactic plane consistently with WMAP. Instead of introducing random variations in the parameters, as we do in GP, we rotate the full-sky map by a random angle before extracting the sky-patch shown in Fig.~\ref{mask}. In this approach, the pixel-to-pixel variations are larger than those obtained with the GP realizations. Since each rotation angle brings a different region of the Galactic template into the observed patch, the realizations differ not only in their pixel values but also in their spatial structure, whereas GP and d11s6 always reuse the same sky region. By construction, RP is therefore the most morphologically diverse of the foreground models considered in this work. We characterize this diversity quantitatively in Section~\ref{morph_div}, where we use the spectral variability of the models, measured through the dispersion of their $B$-mode power spectra across realizations, together with the marginal statistics. To avoid the inclusion of the Galactic plane in the observed patch, we set an arbitrary threshold of $40\,\mu$K and reject any rotation angle for which the maximum value of $Q$ at 220~GHz, within the HEALPix base pixel that contains the observed patch, exceeds this threshold. This criterion is applied because the foreground intensity in these regions is orders of magnitude larger.

\subsubsection{PySM3 d11s6}

This foreground model corresponds to the d11s6 model of PySM3, and the variations are introduced by the library itself. For both dust and synchrotron emissions, this method introduces stochastic small-scale fluctuations generated on the fly on top of the fixed d10 and s5 templates. The d10s5 model is based on templates from the GNILC needlet-based analysis of Planck data, with reduced contamination from the cosmic infrared background and point sources compared to the Commander maps used in d1s1\footnote{\url{https://pysm3.readthedocs.io/en/latest/models.html}}.

\subsubsection{PySM3 d4s2}

This model generalizes the dust model d1 to multiple dust populations using the two-component model described in \cite{finkbeiner1999extrapolation}. The synchrotron model s2 is the same as the one we used in RP. We include this model among the testing options for comparison with \cite{yan2024cmbfscnn}.

\subsubsection{PySM3 d12s7}

The dust contribution in this model relies on a three-dimensional model of polarized dust emission with six layers, based on \cite{2018MNRAS.476.1310M}. The synchrotron contribution consists of a power-law with a frequency-dependent spectral index. This synchrotron model uses the same templates and spectral index map as s5, and includes a curvature term derived from the ARCADE experiment \cite{kogut2012synchrotron}, along with additional small-scale random fluctuations. We include this model among the testing options, following \cite{bianchini2025cmb}.

\section{Foreground cleaning methods}\label{methods}

In this section, we present our procedure to remove Galactic foregrounds from CMB polarization maps. For comparison, we also perform a maximum-likelihood parametric approach and a version of the Internal Linear Combination (ILC) method applied to CMB polarization.

\subsection{Maximum likelihood parametric method}\label{parametricmet}

This method assumes a frequency-dependent model for the Galactic foregrounds and the instrument. This method is implemented using the FGbuster library\footnote{\url{https://github.com/fgbuster/fgbuster}} \cite{Stompor_2009}. We use the frequency-dependent, contaminated, and noisy maps as inputs, with the 85–270 GHz bands brought to a common beam of 25.5 arcmin, i.e. the same inputs as the ILC (Section~\ref{ILCmet}), since the pixel-domain fit assumes a single angular resolution, and the algorithm recovers the CMB, dust, and synchrotron maps. It is important to note that this method relies on estimating the parameters of a given model (e.g., Eq.~\ref{MBB} for dust), and therefore its performance is sensitive to the assumed frequency model. Here, we first estimate the spectral parameters and then solve a linear system of equations to obtain the amplitudes at each pixel. 

\subsection{Internal Linear Combination (ILC) method}\label{ILCmet}

The ILC is a method that does not assume a specific foreground model, although it does assume that the CMB is the same for all frequencies, which holds in $\mu$K$_{\rm CMB}$ units if the CMB energy density corresponds to a blackbody spectrum. It also assumes linearity and minimum variance to build the estimator. In this paper, we use a pixel-domain polarization ILC adapted for polarization maps, which operates on the $Q$ and $U$ maps \cite{Fern_ndez_Cobos_2016,zhang2022efficient}. In our model, we assume that

\begin{equation}\label{ILC}
\begin{split}
\hat{Q}_{\mathrm{CMB}}(p) \pm i\,\hat{U}_{\mathrm{CMB}}(p)
&= \sum_{j=1}^{N_\nu}
\left(
\omega_j^{(R)} \pm i\,\omega_j^{(I)}
\right) \\
&\qquad \times
\left[
Q_j(p) \pm i\,U_j(p)
\right],
\end{split}
\end{equation}
where $p$ represent the pixel in the sky, $N_\nu$ is the number of frequency bands, $Q_j(p)$ and $U_j(p)$ are the beam-convolved, frequency-dependent contaminated and noisy maps, and $\hat{Q}_{\mathrm{CMB}}(p)$ and $\hat{U}_{\mathrm{CMB}}(p)$ are the CMB maps estimated by the method. This technique relies on estimating the real and imaginary part of the weights $\omega_j^{(R)}$ and $\omega_j^{(I)}$. The weights are obtained by minimizing the variance 
\begin{equation}
    \left<\left|\hat{Q}_{\rm CMB} + i\,\hat{U}_{\rm CMB}\right|^2\right>
\end{equation}
which is the only relevant quantity under the assumption that the CMB is Gaussian \cite{mccarthy2025signal}.

Once the weights are estimated, the predicted cleaned map is obtained as the linear combination defined in Eq.~\ref{ILC}. An important aspect of this method is that all foreground-contaminated maps must have the same angular resolution. Since it is not possible to upgrade a map's resolution, we degrade all frequency maps to the lowest resolution among the frequency bands. We find that the resolution of the 30 and 40~GHz channels is too low to obtain reliable ILC results. Therefore, to apply this method, we convolve all frequency maps to a common beam size of 25.5~arcmin and exclude the 30 and 40 GHz bands from the analysis. In Appendix A, we show the difference between the ILC performance applied to the d1s2 model with and without the 30 and 40 GHz frequency bands.

\subsection{CNN-based method}

For the traditional methods described above, it is important to note that more recent and advanced techniques exist. Examples include MCMC-based estimates of spectral parameters rather than relying on the maximum-likelihood solution, as well as more complex and realistic models of the likelihood. There are also more sophisticated ILC approaches, such as the Harmonic ILC (HILC) \cite{kim2009cmb}, the Needlet ILC (NILC) \cite{basak2013needlet}, among others. The traditional methods presented in this work should therefore be understood as proof-of-concept approaches for comparison with the CNN-based method. Consequently, we do not claim that the CNN-based method outperforms state-of-the-art traditional techniques, since we are not comparing against the most recent and optimized algorithms. This is not the point of this paper.

However, both traditional methods discussed before and their upgraded versions present significant implementation challenges. The parametric approach assumes a specific FM and then estimates its parameters. As a result, any inaccuracy in the assumed model propagates directly into the reconstructed maps. The ILC method does not suffer from this limitation because it is blind to the FM. Both methods, however, operate on input maps brought to a common angular resolution, which entails a loss of information. Since it is not possible to increase the resolution of a map, all maps must be degraded to the resolution of the lowest-frequency channel. This process inevitably leads to a loss of small-scale information contained in the higher-resolution maps.

Motivated by these challenges, we introduce a CNN-based method that assumes a foreground model during both training and testing. During training, the CNN is expected to learn the most relevant features for removing the Galactic foregrounds. Then, the trained algorithm can be applied to data simulated from a different FM to assess its generalization capabilities and estimate the error. This method could be applied to real data if the test set contained a feasible physical model of the foregrounds and generalization was achieved, both of which depend strongly on the training set. An advantage of this approach is that it does not require degrading the angular resolution of the input maps. Instead, the CNN is fed with the $Q$ and $U$ frequency maps at each frequency band's original resolution, allowing it to use all the information in the input maps. 

The CNN-based method consists of training a neural network that takes as input the foreground-contaminated and noisy polarization maps and produces the corresponding noisy CMB map at the target frequency, which in our case is $220\,\mathrm{GHz}$. We choose this output frequency based on the case-1 experiment shown in Table 3 of \cite{yan2024cmbfscnn}, where using 220 GHz as the output frequency is found to give results similar to 155 GHz. Although the 155 GHz output band performs slightly better, we adopt 220 GHz in this work in order to probe smaller scales, taking advantage of the higher angular resolution (smaller beam) of this channel. In the following sections, we describe the relevant details of the method.

\subsubsection{Data pre-processing}\label{prepro}

In Section~\ref{sims}, we describe the procedure used to generate partial-sky simulations including CMB, Galactic foregrounds, and instrumental noise. In this section, we describe how we transform the Healpy sky maps into the input of the CNN-based method. Since CNNs operate on images, the $12 \times N_{\rm side}^2$-dimensional Healpy sky maps must be converted into two-dimensional blocks with shape $(N_{\rm size}, N_{\rm size})$, where $N_{\rm size}$ refers to the number of pixels of the block and is not necessarily equal to $N_{\rm side}$. This transformation is applied to both $Q$ and $U$ maps for each frequency band independently. To perform this conversion, we follow the technique described in \cite{yan2023recovering}. This operation is not a projection; instead, it is a bijective mapping in which the pixels belonging to each $N_{\rm side}^2$ sky patch are reorganized one-to-one into a $(N_{\rm side}, N_{\rm side})$ block.

The previously mentioned bijective mapping is done preserving the local structure of the map through the Morton curve re-indexation. First, we consider the $N_{\rm side}$ pixels included in a given Healpix face. These pixels are organized in a 1-dimensional array. Then, we apply the Morton re-indexation \cite{morton1966computer} to assign a 2-dimensional index to each pixel. By construction, this procedure preserves the local structure of the sky map at the block level. The local structure is only broken if we consider different faces of the Healpy map, which is not done in this work.

Due to the limited sky area covered by the observed patch, several pixels within the corresponding sky patch (with size ($N_{\rm side}$, $N_{\rm side}$)) are fixed to zero. This allows us to reorganize the pixels extracted from a sky with $N_{\rm side}=256$ into a block with shape $(N_{\rm size}=128,N_{\rm size}=128)$. Since CNN training is computationally expensive, reducing the input size is advantageous, as it enables larger batch sizes and reduces overall training time.

Finally, once the foreground-contaminated and noisy blocks are constructed, we compute and store the maximum absolute pixel value across all channels and normalize the input blocks by this factor. This normalization ensures that all input channels lie in the range $[-1,1]$. Since the normalization factor is saved, the CNN output can later be rescaled to compute the angular power spectra.

\subsubsection{Training of the CNNs}

The training procedure of the CNN-based method begins with selecting a network architecture, followed by generating training and validation sets of $1000$ and $300$ foreground-contaminated blocks with added noise, respectively, as described in Section~\ref{prepro}. The CNN takes inputs with shape $(N_{\rm batch}, N_{\rm channels}, N_{\rm size}, N_{\rm size})$. In this work, we use $N_{\rm batch}=32$ and $N_{\rm channels}=16$, corresponding to eight frequency bands for each of the $Q$ and $U$ polarization maps. As discussed in Section~\ref{prepro}, the image size is $N_{\rm size}=128$ for simulations with $N_{\rm side}=256$.

The output of the CNN is a tensor with shape $(2,N_{\rm size},N_{\rm size})$, corresponding to the reconstructed noisy CMB $Q$ and $U$ blocks. The training is performed over $60{,}000$ iterations. At each iteration, $N_{\rm batch}$ foreground-contaminated blocks are randomly selected, stacked, and fed to the CNN. The network processes the input tensor, and the loss function consists of the Mean Absolute Error (MAE), computed by comparing the CNN output with the target noisy CMB $Q$ and $U$ blocks pixel-to-pixel. In addition, we compute the Fast Fourier Transform (FFT) of both the output and target blocks, and include this information in the loss function as a physics-guided term \cite{yan2024cmbfscnn}. The FFT-term computes the MAE in the same way as before, but applied to the amplitude of the Fourier-transformed image

\begin{equation}
    {\rm A_F(I) = \sqrt{Re[FFT(I)]^2+Im[FFT(I)]^2}}.
\end{equation}

We adopt MAE rather than other metrics like the MSE in pixel space because it is less sensitive to the small number of high-amplitude pixels — such as those near the brightest foreground regions — which would otherwise dominate a quadratic loss, and because it tends to produce less over-smoothed reconstructions; this choice also follows \cite{yan2024cmbfscnn}. The total loss function used during training is therefore identical to the one defined in \cite{yan2023recovering} and is given by
\begin{equation}
    \mathcal{L} = \mathcal{L}_{\rm MAE} + \mathcal{L}_{\rm FFT}.
\end{equation}

Gradients are computed using PyTorch, and backpropagation is used to update the CNN's trainable parameters. For all trainings, the reported network corresponds to the iteration of minimum validation loss (checkpoint selected via early stopping on the validation set~\cite{goodfellow2016}), rather than the final training iteration. While a persistent train/validation gap is observed, especially for the networks trained on RP, this selection criterion ensures that all quoted generalization results are based on the best-generalizing checkpoint available, not an overfit endpoint. The training and validation curves are shown in Appendix B.

\subsubsection{CNN architectures}\label{comp_section}

In this work, we consider two different CNN architectures. The first one is a UNet-type architecture (UN) with two convolutional layers per level\footnote{\url{https://github.com/jaxony/unet-pytorch/}} \cite{ronneberger2015unetconvolutionalnetworksbiomedical}. The second architecture is a more complex model presented in \cite{yan2023recovering}, referred to here as L3\footnote{\url{https://github.com/yanyepeng/CMBFSCNN/blob/main/cmbfscnn/CNN_models.py}}. The L3 architecture divides the input images into multiple blocks, processes each block with internal UNets, and subsequently applies convolutional layers with batch normalization and dilated convolutions to obtain the final output image. In this work, we modified the original L3 architecture to produce two output channels, corresponding to the $Q$ and $U$ polarization components, because the original design has only one output channel \cite{yan2023recovering}.

The training time for both architectures is comparable, totaling approximately $8$ hours on an A100 GPU. UN contains approximately $1.2\times10^8$ parameters, whereas the L3 architecture contains approximately $8\times10^6$ parameters. Despite its smaller parameter count, the L3 architecture's increased computational complexity results in training time comparable to the other architecture.

\subsubsection{Testing the CNNs}

After the training stage, we evaluate the performance of the CNN-based method on a test set of $300$ maps generated using the same procedure described in Section~\ref{prepro}. These maps are completely independent of those used during training and validation. Each testing map is processed individually: the foreground-contaminated and noisy blocks are fed through the trained CNN to obtain the predicted noisy CMB blocks. Then, we apply the inverse transformation to recover the partial-sky maps from the output blocks. After that, we compute the angular power spectra of both the target and reconstructed maps using the pseudo-$C_\ell$ technique implemented in NaMaster \cite{alonso2019unified}. The performance of the CNN is quantified using the ratio $C_\ell^{\rm Rec, X} / C_\ell^{\rm True, X}$, where $X=E,B$.

\section{Generalization hypothesis}\label{gen}

The goal of this work is to evaluate how training-model assumptions impact CNN performance in CMB polarization foreground cleaning. In this section, we introduce a set of pixel-to-pixel statistical quantities to characterize which properties of the training data are most relevant for enabling generalization. These statistics quantify the marginal variability and complexity of the training distribution and provide a framework to interpret the empirical generalization performance of the CNNs studied in this paper. In this work, we do not attempt to measure learning difficulty directly. Instead, we use the marginal and spectral statistics defined below as complementary measures of the breadth of the training distribution, that is, of the range of pixel values and of spatial structures to which the CNN is exposed during training. Statements about a model being more ``complex" should be read in this sense throughout, and not as a claim about physical complexity or about other notions such as the number of layers or the degree of non-Gaussianity. To simplify the discussion, we will focus on the polarization amplitude $P = \sqrt{Q^2 + U^2}$.
This choice captures the overall amplitude of the polarized emission and allows for a direct comparison between different FMs.

We use a full-sky frequency-dependent FM model, and apply the 2D image mapping described in Section \ref{prepro} to generate realizations of the polarization maps $Q$ and $U$, with dimensions $[N_{\rm freqs}, N_{\rm side}, N_{\rm side}]$. The amplitude of $P$ at each pixel will be denoted as $\mathcal{P}_{ijk}^{Mr}$, where $i \in [1, N_{\rm freqs}]$ labels the frequency channel, $j,k \in [1, N_{\rm side}]$ label the pixel indices, $M$ refers to the specific model assumed, and $r$ indexes the realization number. Throughout this analysis, we consider a single-frequency channel, fixed at $i$, corresponding to $220$~GHz. The statistics of this section and the spectra of
Figs.~\ref{cls_foreg} and \ref{cls_foreg_wide} are computed from maps  at $N_{\rm side}=128$.

\subsection{Foreground models marginal statistics}\label{statistics}

For a given FM ($M$), channel ($i$), and pixel $(j,k)$, we define the mean over $R$ realizations as
\begin{equation}
    \mu[P_{ijk}^M] = \frac{1}{R} \sum_{r=1}^R P_{ijk}^{Mr}.
\end{equation}
Using this definition, the unbiased variance is given by
\begin{equation}
    {\rm Var}[P_{ijk}^M] = \frac{1}{R-1} \sum_{r=1}^R \left(P_{ijk}^{Mr} - \mu[P_{ijk}^M]\right)^2.
\end{equation}

The mean and variance correspond to the first- and second-order moments of the distribution. For a Gaussian distribution, these quantities would fully characterize its marginal statistical properties. However, since the FMs considered in this work are non-Gaussian, we introduce two additional statistics: the skewness and the Shannon entropy.

The skewness measures the asymmetry of the distribution around its mean and vanishes for a Gaussian distribution. We compute the skewness at a given pixel as
\begin{equation}
    {\rm Skew}[P_{ijk}^M] = \sqrt{R}
    \frac{\sum_{r=1}^R \left(P_{ijk}^{Mr} - \mu[P_{ijk}^M]\right)^3}
    {\left(\sum_{r=1}^R \left(P_{ijk}^{Mr} - \mu[P_{ijk}^M]\right)^2\right)^{3/2}}.
\end{equation}

The final quantity used to characterize the FMs is the Shannon entropy \cite{shannon1948mathematical}. Unlike moment-based statistics, the Shannon entropy probes the effective support of the probability distribution and provides a measure of how predictable the polarization amplitude is across realizations. In this work, we estimate the Shannon entropy using a simple binning approximation. For a fixed channel and pixel, the values of $P_{ijk}^{Mr}$ across the $R$ realizations are binned to obtain an empirical probability distribution $p_{ijk}^M(b)$. Here, b represents the index of the histogram bins used to discretize the polarization amplitude values across realizations. Note that the binning scheme is the same for all the pixels. We adopt $R = 1000$ realizations and $N_{\rm bins} = 100$ bins. We adopt this binning scheme because using a larger number of realizations just makes the computation slower, and we confirmed that it does not change the results. The Shannon entropy is then defined as
\begin{equation}
    S[P_{ijk}^M] = -\sum_{b=1}^{N_{\rm bins}} p_{ijk}^M(b) \, \ln p_{ijk}^M(b).
\end{equation}

This computation is a first-order approximation because it neglects correlations between frequency channels and between neighboring pixels. Nevertheless, for a fixed channel and pixel, it provides a useful indicator of the foreground signal's variability across realizations. This is particularly informative in template-based simulations, where large fractions of the sky exhibit vanishing or weak foreground emission and are therefore easier to predict than regions with large foreground amplitudes. An additional useful property of the Shannon entropy is that it is bounded from above by the value corresponding to a uniform distribution. For $N_{\rm bins} = 100$, this maximum value is $\ln(100) \approx 4.6$. Moreover, this quantity is also bounded below by $S=0$, reaching this value when the probability distribution corresponds to a Kronecker delta. A small value of $S$ implies a predictable value of the pixel, while a larger one implies a harder prediction of the pixel, but generated by a uniform distribution, which is not what the physics of the foregrounds generates.


\subsection{Marginal complexity and morphological diversity}\label{morph_div}

In the previous section, we presented some statistics to quantify the marginal complexity for a given frequency channel. These quantities give us information regarding the change in the polarization amplitude $P$ at each pixel through realizations, but they are not the only relevant properties of the models. For instance, if we followed the criteria of maximizing the marginal statistics, a possible model to train the CNNs could be a random uniform sampling for each pixel. This model could maximize the variance as well as the Shannon entropy. However, we would not expect that training under a random sampling to drive generalization to the CNN-based model. A deeper discussion about this is presented in Section~\ref{GP_comp_sec}, where we show that a GP modification whose SED parameters are sampled independently in each pixel (GP-W8M) attains marginal statistics comparable to GP-W8 but yields a reconstruction whose noise properties are not controlled, showing that comparable marginal statistics do not guarantee a usable reconstruction. Under this situation, we impose a restriction on our FMs, which is always true for template-based models: all the realizations have to follow the spatial structure of the measured Galactic foregrounds. For template-based models, this is always true because the intensity distribution is based on observations. Then, an important consequence emerges: the performance of the CNN-based method is restricted to this shared spatial distribution.

We will define the morphology diversity as the variability of spatial structures through the realizations of the model. One must note that it is not easy to produce an arbitrarily large number of morphologically diverse realizations for template-based models. However, we can generate some of them by splitting an observational template into different patches and then using them as different realizations. Following this idea, the RP model emerges as the most morphologically diverse model, while by construction, GP and d11s6 generate new realizations from a fixed patch of the observational template.

A relationship emerges between marginal complexity and morphological diversity when the models under consideration share the same morphology, as observed in the Shannon entropy. This quantity measures how easy it is to predict the value of $P$ for a given pixel. If the model is based on a fixed template (like GP or d11s6), the value of a given pixel is easier to predict compared to models like RP. Indeed, RP has no pixels consistent with $S=0$, unlike GP or d11s6 (see Fig.~\ref{metricsP}). We establish a relationship between this result and the morphological diversity: if all the models followed the same morphology, the unpredictability of the pixels would be a measure of the diversity of the model. We remark that our calculation of the Shannon entropy is just a zeroth-order approximation. A full calculation of the Shannon entropy, including correlations between pixels and frequency channels, would also give us information about the morphology diversity and frequency dependence. However, the numerical calculation of this quantity is computationally expensive, and we leave this for future work.

\begin{figure}
    \centering
    \includegraphics[width=\linewidth]{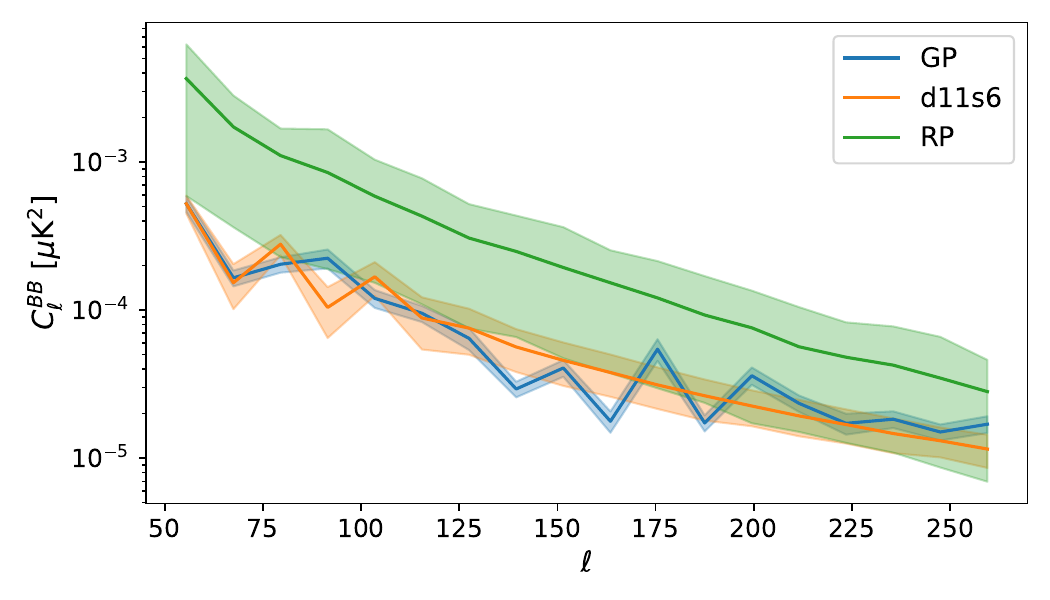}
    \caption{Average and 68\% probability interval of the $C_\ell^{BB}$ computed for GP, d11s6 and RP using 1000 realizations. This figure shows the spectral variability at the level of the two-point correlation function.}
    \label{cls_foreg}
\end{figure}

A simple calculation of the B-mode power spectra for the foreground models considered in this work quantifies the spectral variability of the models at the level of the two-point correlation function. This quantity is not a measure of morphological diversity alone, since it also responds to changes in the overall foreground amplitude; we therefore use it as a second measure of the breadth of the training distribution, complementary to the marginal statistics. In Figure \ref{cls_foreg}, we show the averaged $C_\ell^{BB}$ and the 68\% probability interval (PI), between the 16th and 84th
percentiles, computed over 1000 realizations for each model. This quantity summarizes the previous discussion: RP has the widest band, d11s6 follows it, and GP has the narrowest band. At the level of the two-point function, RP therefore has the
largest spectral variability, followed by d11s6, consistent with RP being
the most morphologically diverse model by construction. It is interesting to note that the RP averaged values are smoothed at all multipole scales shown here because of the wide range of variations that this model has. Because d11s6 introduces variations at a small scale, this smoothing is happening only for $\ell>100$. As GP only rescales the template globally, the corresponding power spectra are not smoothed for any value of $\ell$.

We also computed the average over the range $50<\ell<260$ of the half-width of the 68\% PI divided by the mean. This quantity gives 0.14 for GP, 0.28 for d11s6 and 0.76 for RP. This result follows the previous discussion and quantifies the spectral variability of each model at the two-point correlation function level.

\begin{figure}
    \centering
    \includegraphics[width=\linewidth]{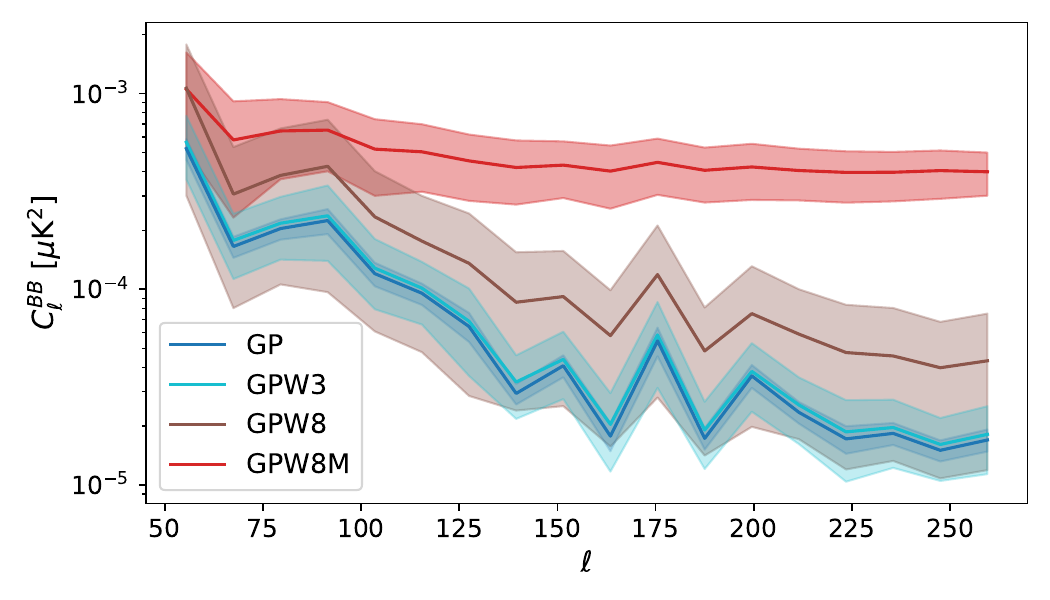}
    \caption{Average and 68\% probability interval of the $C_\ell^{BB}$ computed for the modifications of GP using 1000 realizations. This figure shows the spectral variability at the level of the two-point correlation function.}
    \label{cls_foreg_wide}
\end{figure}

In Figure~\ref{cls_foreg_wide}, we repeat this calculation for the GP modifications. Averaged over $50<\ell<260$, the relative width of the band grows from $0.14$ for GP to $0.39$ for GP-W3 and $0.73$ for GP-W8, while the shape of the mean spectrum is essentially preserved. GP-W8 is therefore the modification that maximizes the spectral variability while retaining the morphology inherited from the template. GP-W8M behaves differently: since its SED parameters are sampled independently in each pixel, the spatial coherence of the template is erased by construction, and the mean spectrum becomes nearly flat, resembling white noise. Its band is in fact narrower ($0.36$) than that of GP-W8, despite both models having comparable marginal statistics. A broad marginal distribution therefore does not by itself imply a diverse set of spatial structures.

\subsection{Analysis of foreground models}\label{hierarchy}

\begin{figure*}[ht!]
    \centering
    \begin{minipage}{\textwidth}
        \centering
        \includegraphics[width=\textwidth,height=0.28\textheight,keepaspectratio]{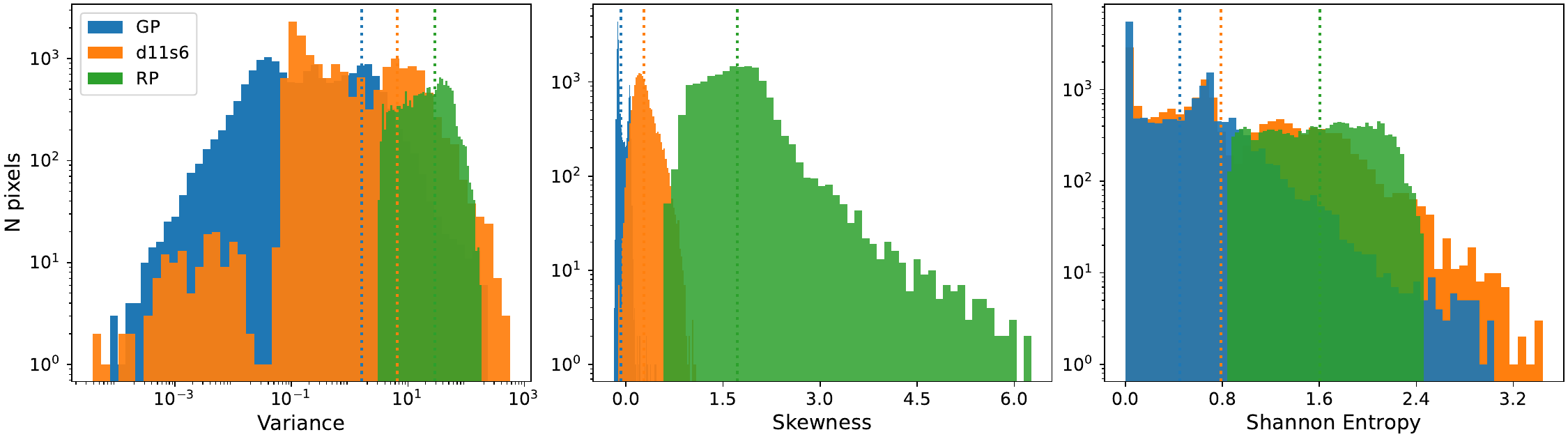}
    \end{minipage}

    \begin{minipage}{\textwidth}
        \centering
        \includegraphics[width=\textwidth,height=0.28\textheight,keepaspectratio]{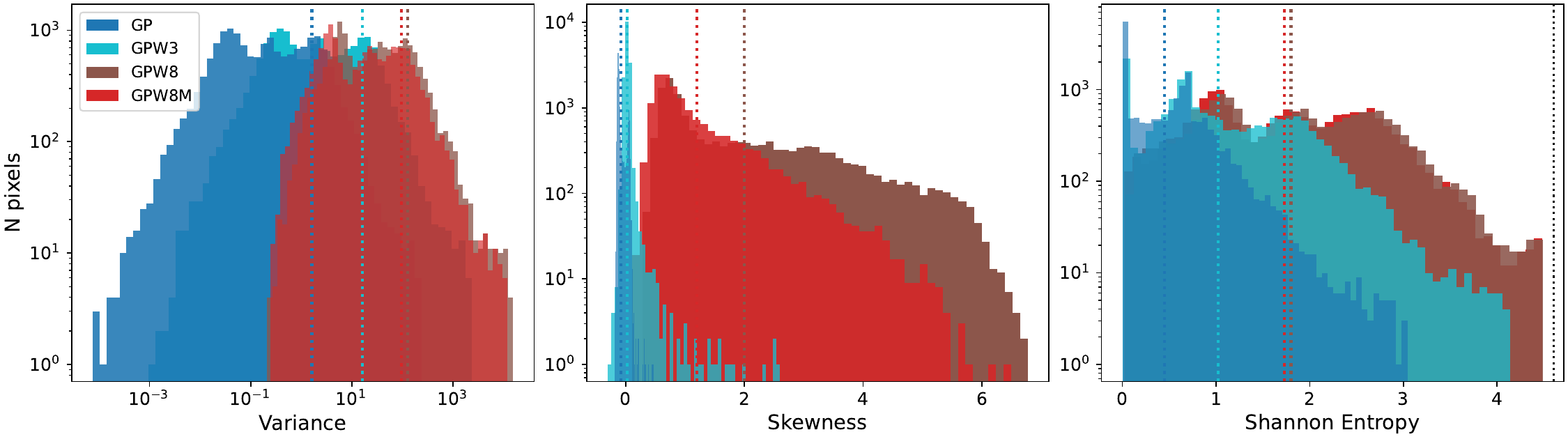}
    \end{minipage}

    \caption{Pixel-level distributions of the statistical metrics computed for the polarization amplitude $P=\sqrt{Q^2+U^2}$ of the foreground models used for the training. 
    The vertical colored dotted lines indicate the mean value over all pixels in the analyzed block, which are reported in Table~\ref{metrics_num}; the black dotted line in the lower-right panel marks the maximum entropy $\ln(100)\approx4.6$ allowed by our binning. The upper panel corresponds to GP, d11s6 and RP, while the lower panel to the modifications of GP.}
    \label{metricsP}
\end{figure*}

The statistics described in the section \ref{statistics} are shown in the upper panel of Fig.~\ref{metricsP} for GP, d11s6 and RP. Since the pixel indices $ j$ and $k$ are not fixed, each pixel in the analyzed block contributes a single value to the statistics. The resulting pixel-level values are therefore displayed as histograms.

The Shannon entropy distributions for the d11s6 and GP models show a significant fraction of pixels falling into bins consistent with $S \approx 0$. This behavior indicates that, for these models, certain pixels are highly predictable across realizations. In contrast, the entropy distribution of the RP model exhibits a nonzero lower bound, implying that no individual pixel is trivially predictable in this case.
The skewness distributions further show that RP has the strongest asymmetry with respect to the mean. 
Finally, in terms of the variance, d11s6 and GP show less variability than the RP model. The mean values of all the metrics are summarized in Table~\ref{metrics_num}.

We define the marginal statistical complexity of a foreground model as the degree of pixel-level variability, asymmetry, and unpredictability across realizations, as quantified by the variance, skewness, and Shannon entropy, respectively. According to these criteria, RP is identified as the most statistically complex FM at the pixel-level, while d11s6 and GP exhibit a similar statistical behavior between each other. Here we also note the discussion in the previous section, where we concluded that RP is also the most morphologically diverse model.

\begin{table}[h!]
    \centering
    \begin{tabular}{|c|c|c|c|}
        \hline
        $P$ (220~GHz) & Variance [$\mu$K$^2$] & Skewness & Entropy \\
        \hline\hline
        GP      & 1.6 & $-0.08$ & 0.4 \\
        \hline
        GP-W3   & 16  & 0.02    & 1.0 \\
        \hline
        GP-W8   & 125 & 2.0     & 1.8 \\
        \hline
        GP-W8M  & 96  & 1.2     & 1.7 \\
        \hline\hline
        d11s6   & 6.5 & 0.3     & 0.8 \\
        \hline
        RP      & 29  & 1.7     & 1.6 \\
        \hline
    \end{tabular}
    \caption{Mean values of the statistical metrics considered in this work (shown as vertical dotted lines in Fig.~\ref{metricsP}). The Shannon entropy is bounded by $\ln(100)\approx4.6$ for the binning adopted here.}
    \label{metrics_num}
\end{table}

Based on these results we formulate the following generalization hypothesis: CNNs trained on the RP model are expected to exhibit an improved performance if we test on template-based FMs different from those used during training, compared to CNNs trained on d11s6 or GP. This generalization performance is computed with the performance metrics discussed in the following section. 

The lower panel of Fig.~\ref{metricsP} shows the same statistics for the GP
modifications, whose mean values are listed in Table~\ref{metrics_num}. The
three metrics grow from GP to GP-W3 and GP-W8. Compared with RP,
GP-W3 remains below in all three of them, whereas GP-W8 exceeds RP in all of
them. In addition, the entropy distributions of GP-W8 and GP-W8M extend up to
$\approx4.5$, close to the maximum value $\ln(100)\approx4.6$ allowed by our
binning, indicating that for these models a fraction of the pixels becomes
essentially unpredictable across realizations. GP-W8M shows marginal statistics
comparable to those of GP-W8, as expected, since the two models share the
amplitude of the variations and differ only in whether each parameter is sampled
once per map or independently in each pixel.

\subsection{Performance metrics}

In the previous section, we characterized the statistical properties of the FMs by computing the variance, skewness, and Shannon entropy, and used these quantities to establish a hierarchy of complexity among them. We now introduce a set of performance metrics designed to quantify the generalization capability of the CNN-based foreground–removal method. Our goal is to assess how accurately and precisely the reconstructed CMB maps reproduce the true CMB statistics when the foreground distribution differs from that seen during training. To this end, we compare the angular power spectra of the reconstructed and true CMB maps through the following dimensionless ratio, evaluated during the testing stage:
\begin{equation}\label{ratio}
    {\rm rat}_\ell^{X,r}{\rm [M_1,M_2]} =
    \frac{C_\ell^{{\rm Rec},X,r}{\rm [M_1,M_2]}}{C_\ell^{{\rm True},X,r}{\rm [M_2]}} ,
\end{equation}
where $X$ denotes either $E$- or $B$-mode polarization, $r$ labels realizations in the testing set, and the notation ${\rm [M_1,M_2]}$ indicates that the CNN was trained on the FM M$_1$ and tested on FM M$_2$. The set of FMs considered for training is M$_k \in \{\mathrm{GP}, \mathrm{d11s6}, \mathrm{RP}\}$. 

The numerator, $C_\ell^{{\rm Rec},X,r}{\rm [M_1,M_2]}$, is computed from the CNN-reconstructed map and, therefore, depends on both the training and testing models. The denominator, $C_\ell^{{\rm True},X,r}{\rm [M_2]}$, is obtained from the true CMB map associated with model M$_2$ and varies with the realization $r$. In the ideal case of perfect reconstruction, we expect ${\rm rat}_\ell^{X,r}{\rm [M_1,M_2]}$ to be consistent with $1$, for all $\ell$s. To quantify deviations from this ideal behavior, we compute summary statistics of the ${\rm rat}_\ell^{X,r}$ distribution over realizations. As a robust estimator of central tendency, we compute the median
\begin{equation}
    {\rm Med}_\ell^{X}{\rm [M_1,M_2]} =
    {\rm median}\!\left[{\rm rat}_\ell^{X,r}{\rm [M_1,M_2]}\right]_r ,
\end{equation}
which provides a measure of the method's accuracy at each multipole. To characterize the spread of the distribution, we compute the 17th and 83rd percentiles, $p_\ell^{17,X}{\rm [M_1,M_2]}$, $p_\ell^{83,X}{\rm [M_1,M_2]}$, respectively, and with them we define the uncertainty of the method as
\begin{equation}\label{precision}
    \Delta_\ell^X{\rm [M_1,M_2]} = \frac{p_\ell^{83,X}{\rm [M_1,M_2]}-p_\ell^{17,X}{\rm [M_1,M_2]}}{2} ,
\end{equation}
which corresponds to a central 66\% probability interval. We employ medians and percentiles rather than means and standard deviations because the distribution of ${\rm rat}_\ell^{X,r}$ is generally non-Gaussian and asymmetric. The quantity $\Delta_\ell^X$ therefore provides a measure of the reconstruction's precision, quantifying the realization-to-realization scatter. If the method is unbiased, the median satisfies ${\rm Med}_\ell^X{\rm [M_1,M_2]} \approx 1$. Deviations from unity indicate a systematic offset in the reconstructed power spectrum. We therefore define the following estimator:
\begin{equation}\label{accuracy}
    {\rm bias}_\ell^X{\rm [M_1,M_2]} =
    \left| {\rm Med}_\ell^X{\rm [M_1,M_2]} - 1 \right|,
\end{equation}
which quantifies the accuracy of the cleaning method at each multipole. Small values of ${\rm bias}_\ell^X$ indicate that the reconstructed spectrum is consistent with the true one, while larger values show systematic deviations. 

We also compute the number of sigmas that the median deviates from $1$ as a function of $\ell$, combining both previous metrics as

\begin{equation}\label{nsigm}
   {\rm N}\sigma_\ell^X{\rm [M_1,M_2]} = \left| \frac{{\rm Med}_\ell^X{\rm [M_1,M_2]} - 1 }{\Delta_\ell^X{\rm [M_1,M_2]}}\right|.
\end{equation}

These metrics can be straightforwardly applied to other component-separation techniques, such as ILC or parametric methods, by replacing the CNN reconstruction with the corresponding traditional method; for these
methods, $C_\ell^{\rm True}$ is computed from the same noisy 220~GHz  target,
 brought to their common resolution of 25.5 arcmin. While the quantities defined above characterize the performance of a single CNN, we now introduce a metric that directly compares the generalization performance of two networks trained on different FMs. Specifically, we compare CNNs trained on models M$_1$ and M$_2$ when both are tested on a third model M$_3$. We define the bias Generalization Factor (bias-GF) as the difference between the averaged bias estimators,
\begin{equation}\label{sigma_gen_fact}
    \mathcal{G}{\rm bias}_{[M_1,M_2;M_3]}^X =
    \left<{\rm bias}_\ell^X{\rm [M_1,M_3]}\right>
    - \left<{\rm bias}_\ell^X{\rm [M_2,M_3]}\right> ,
\end{equation}
where $\langle \cdot \rangle \equiv \langle \cdot \rangle_{\ell_{\rm min}<\ell<\ell_{\rm max}}$ denotes an average over the multipole range of interest. Analogously, we define a precision-based Generalization Factor ($\Delta$-GF),
\begin{equation}\label{delta_gen_fact}
    \mathcal{G}\Delta_{[M_1,M_2;M_3]}^X =
    \left<\Delta_\ell^X{\rm [M_1,M_3]}\right>
    - \left<\Delta_\ell^X{\rm [M_2,M_3]}\right> .
\end{equation}

In our analysis we adopt $\ell_{\rm min}=50$ and $\ell_{\rm max}=260$. The notation ${\rm [M_1,M_2;M_3]}$ indicates that the CNNs trained on M$_1$ and M$_2$ are compared under identical training conditions, differing only in the FM used to generate the training data, and are both evaluated on the same testing set generated with the FM M$_3$. The sign of the Generalization Factor contains the relevant information: a negative value indicates that the CNN trained on M$_1$ yields smaller deviations from the true CMB spectrum than the one trained on M$_2$ when tested on M$_3$. So, a negative value means that the CNN trained assuming M$_1$ generalizes better than the CNN trained assuming M$_2$, when we test on M$_3$. A positive value indicates the opposite. 
Due to the anti-symmetric property ($\mathcal{G}_{[M_1,M_2;M_3]}^X = -\mathcal{G}_{[M_2,M_1;M_3]}^X$) and the fact that $\mathcal{G}_{[M_k,M_k;M_3]}^X = 0$, only three independent comparisons exist when evaluating three foreground models.
As discussed above, $\mathcal{G}{\rm bias}$ quantifies averaged differences in accuracy, while $\mathcal{G}\Delta$ quantifies averaged differences in precision.

\subsection{Estimation of the tensor-to-scalar ratio}

In this section, we describe the statistical model to compute the tensor-to-scalar ratio, $r$, from foreground-cleaned $C_\ell^{BB}$s. A simple parametrization for the theoretical angular spectra can be assumed as:

\begin{equation}\label{r_param}
    C_\ell^{\rm BB}(r)=r\, C_\ell^{\rm BB,\,{\rm Tens}}(r=1)+C_\ell^{\rm BB,\,{\rm Lens}},
\end{equation}
where $C_\ell^{\rm BB,\,{\rm Tens}}$ is the primordial $B$-mode spectrum with $r=1$ and $C_\ell^{\rm BB,\,{\rm Lens}}$ is the lensing power spectrum. As we are assuming $r=0$ in our simulations,  $C_\ell^{\rm BB,\,{\rm Lens}} = C_\ell^{\rm BB,\,{\rm True}}$. Then, we can build a Gaussian log-likelihood 

\begin{equation}
    -2\ln{\mathcal{L}}(r)=\sum_\ell \frac{\left(\hat{C}_\ell^{\rm BB}-C_\ell^{\rm BB}(r)\right)^2}{\sigma_\ell^2},
\end{equation}
where we assumed a diagonal covariance $\sigma_\ell$. In our case, the foreground-cleaned angular spectra is the estimator, so $\hat{C}_\ell^{\rm BB} = C_\ell^{\rm BB,\, Rec}$. An analytic expression for $\hat{r}$ can be obtained as

\begin{equation}
    \hat{r}=\frac{\sum_\ell \frac{\left(C_\ell^{\rm BB,\, Rec}-C_\ell^{\rm BB,\, True}\right)C_\ell^{\rm BB,\, Tens}}{\sigma_\ell^2}}{\sum_\ell{\left(\frac{C_\ell^{\rm BB,\, Tens}}{\sigma_\ell}\right)^2}}.
\end{equation}

To compute $\sigma_\ell$ we compute the standard deviation of the target power spectra over the realizations of the testing set as

\begin{equation}
    \sigma_\ell^2={\rm Var}\left(C_\ell^{\rm BB, True}\right).
\end{equation}


\section{Results}\label{res}

In this section, we present our results. We adopt the following color code: networks trained on GP are shown in blue, those trained on d11s6 in orange, and those trained on RP in green. Results from the parametric approach are shown in red, while those from the ILC are shown in purple. Throughout this section, we use a compact notation to describe the trained CNNs. We denote by Arch–Train a CNN with architecture ``Arch" trained on a given FM ``Train". For example, L3–GP corresponds to an L3-type network trained on the GP model.

\subsection{Comparison with traditional methods}

\begin{figure*}[ht!]
    \centering

    \begin{minipage}{\textwidth}
        \centering
        \includegraphics[width=\textwidth,height=0.28\textheight,keepaspectratio]{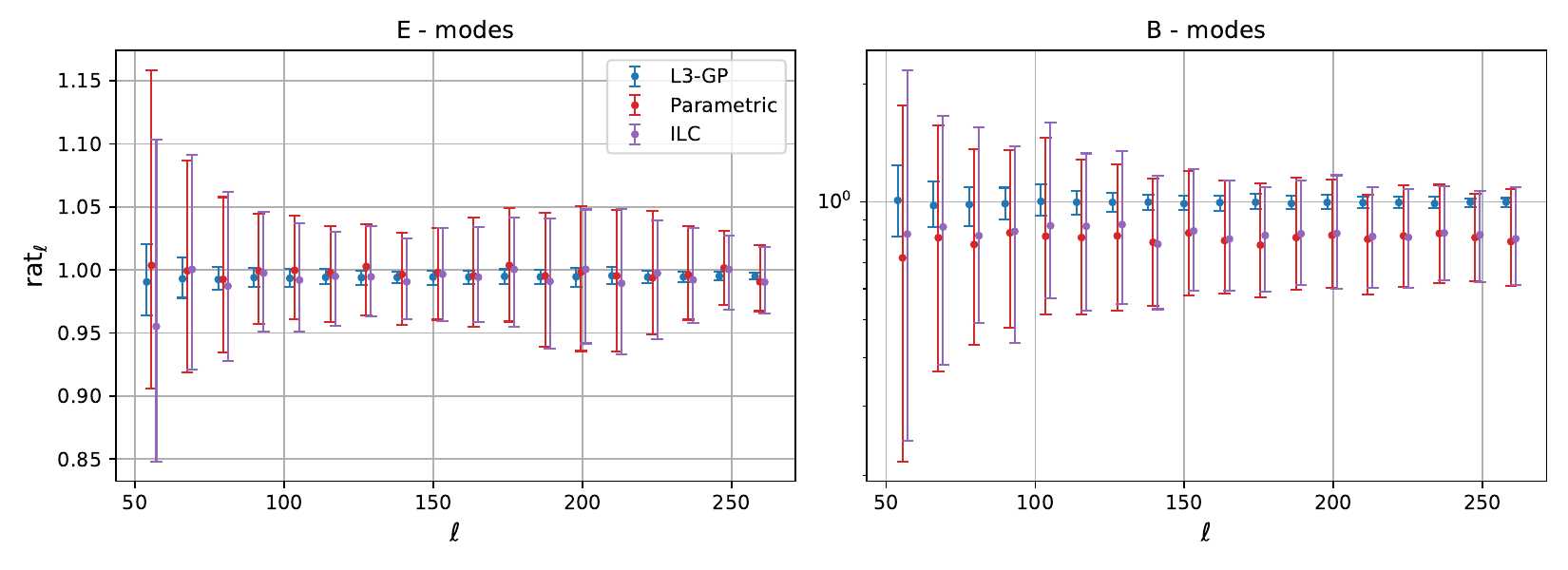}
        \\[-1ex]
        {\small (a) Cleaning methods applied to the Gaussian Parameters (GP) foreground model. For the CNN-based method, we applied the L3 architecture trained on the GP model.}
    \end{minipage}

    \vspace{1ex}

    \begin{minipage}{\textwidth}
        \centering
        \includegraphics[width=\textwidth,height=0.28\textheight,keepaspectratio]{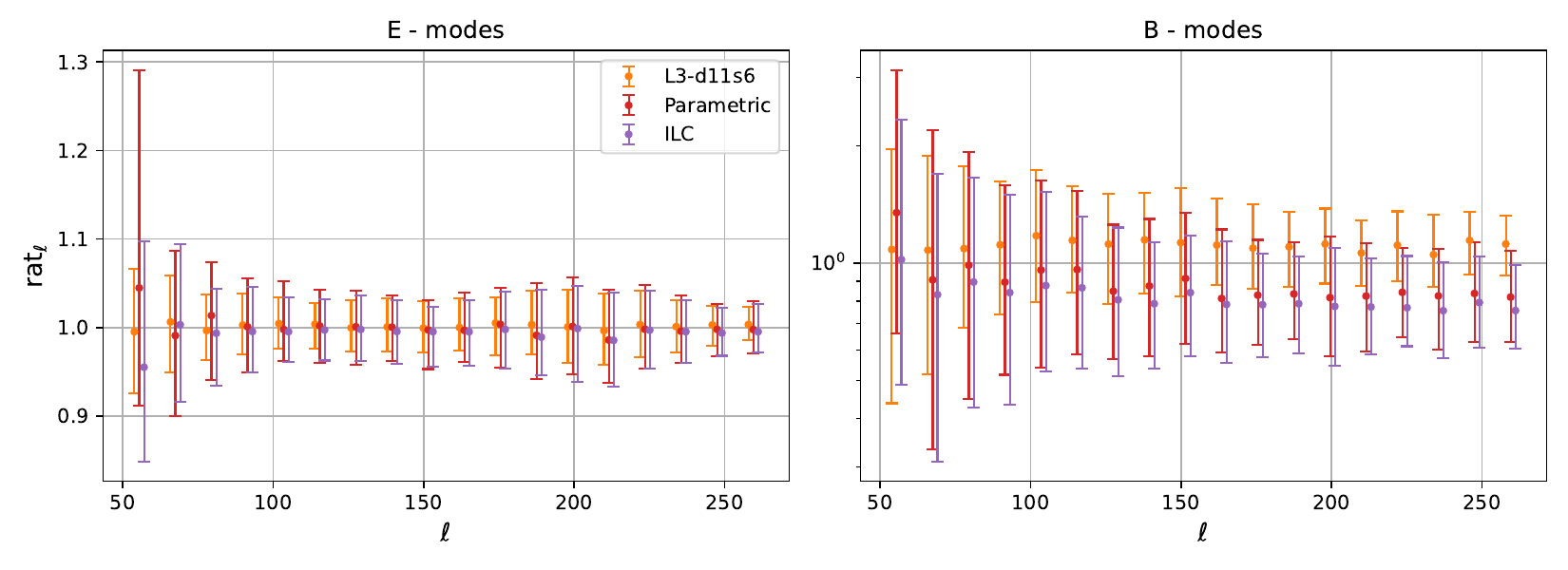}
        \\[-1ex]
        {\small (b) Cleaning methods applied to the PySM3 d11s6 foreground model. For the CNN-based method, we applied the L3 architecture trained on the d11s6 model.}
    \end{minipage}

    \vspace{1ex}

    \begin{minipage}{\textwidth}
        \centering
        \includegraphics[width=\textwidth,height=0.28\textheight,keepaspectratio]{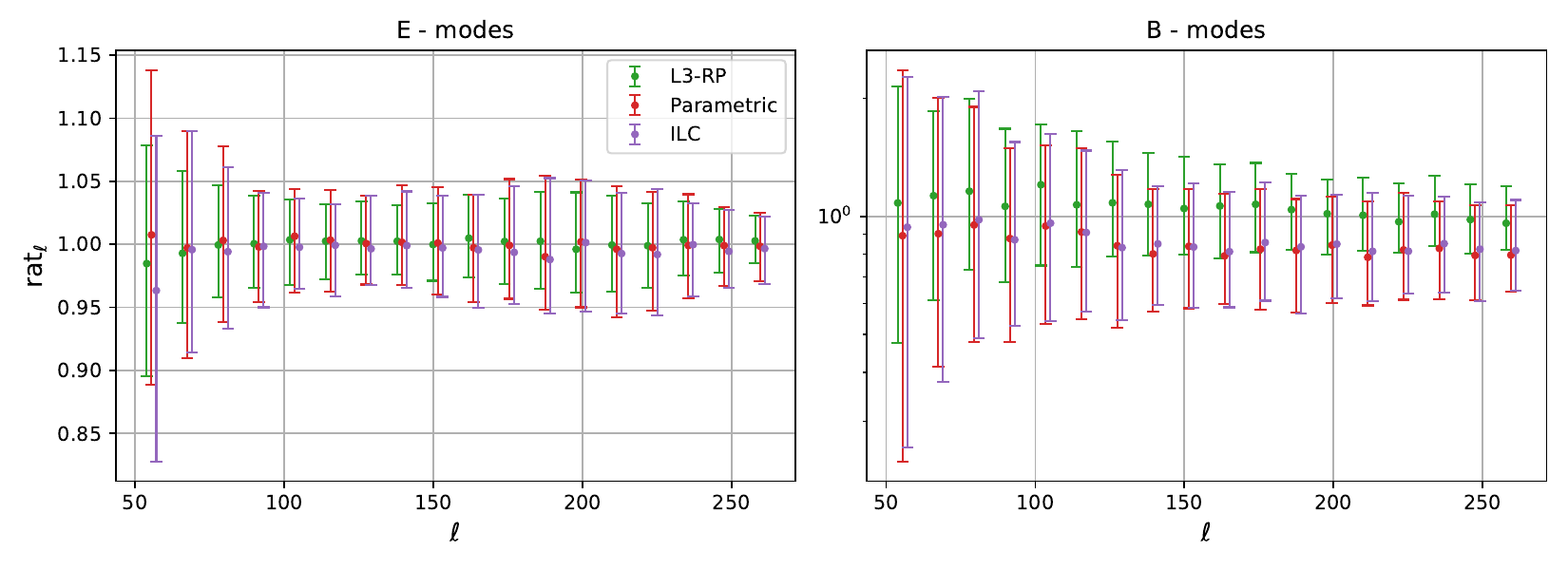}
        \\[-1ex]
        {\small (c) Cleaning methods applied to the Rotating Patches (RP) foreground model. For the CNN-based method, we applied the L3 architecture trained on the RP model.}
    \end{minipage}

    \caption{Median and 66\% Probability Interval (PI) of the ratio defined in Eq.~\ref{ratio}. 
    Left panels correspond to E-modes, while right panels correspond to B-modes. 
    A logarithmic scale is used for the B-modes. 
    We compare the CNN-based method with the traditional methods described in 
    Sections~\ref{parametricmet} and~\ref{ILCmet}.}
    \label{comp_with_trad}
\end{figure*}

In Fig.~\ref{comp_with_trad}, we present the results of foreground removal for the three FMs used to train the CNNs in this work, obtained with the parametric approach (red), the ILC (purple), and the CNN-based method adopting the L3 architecture, due to its specialized design for CMB applications, following the discussion in Section~\ref{comp_section}.

For the three FMs analyzed, the CNN-based method performs better than the traditional methods considered in this work in $B$-modes and in the precision of the $E$-modes, while in $E$-mode accuracy the three methods are comparable, at the $10^{-3}$ level. Moreover, we find that the traditional methods exhibit similar performance across the three FMs used in this analysis. We must be careful when interpreting this result since, in this case, the CNNs were trained and tested on the same FM. This result, therefore, represents a consistency check, as such behavior is expected under these conditions. Since the goal of this paper is to assess the generalization capabilities of the CNN-based method, our main results are based on CNNs trained and tested on different FMs.

\subsection{Qualitative generalization analysis}\label{qualitative_sec}

\begin{figure*}[ht!]
    \centering

    \begin{minipage}{\textwidth}
        \centering
        \includegraphics[width=\textwidth,height=0.28\textheight,keepaspectratio]{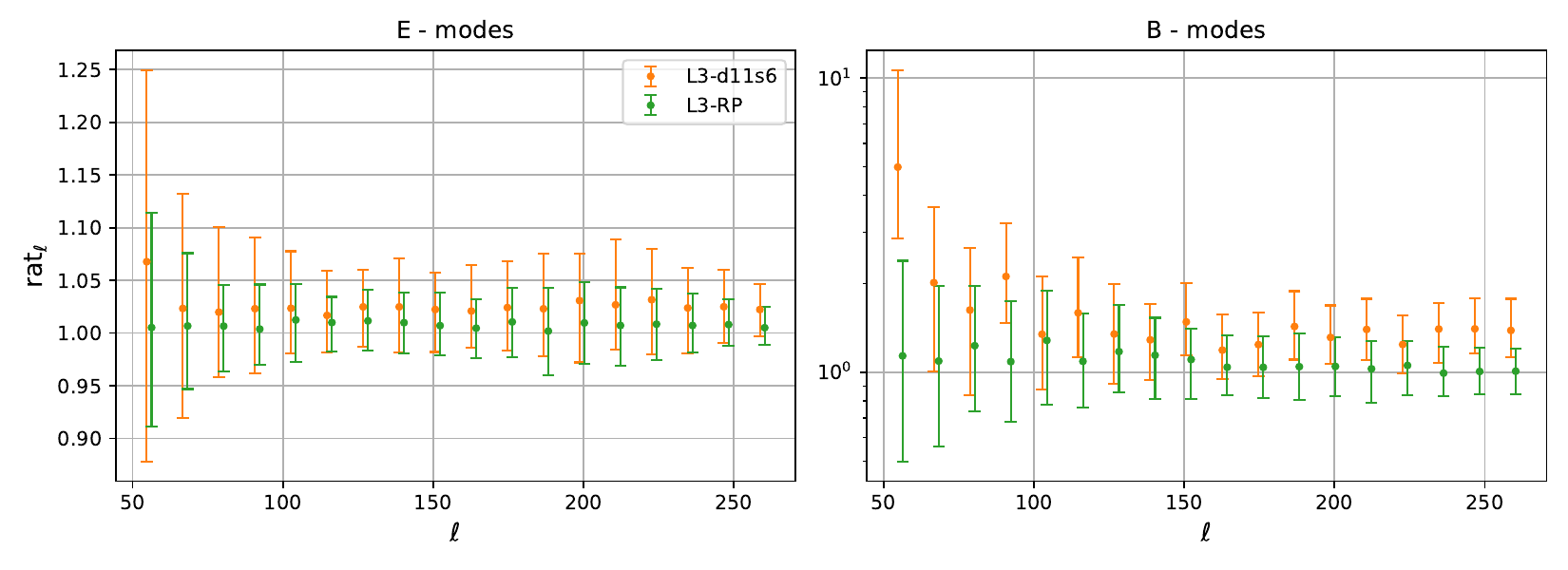}
        \\[-1ex]
        {\small (a) Performance of the CNNs trained on the PySM3 model d11s6 and the Rotating Patches (RP) model and tested on the Gaussian Parameters (GP) model. L3-RP outperforms L3-d11s6 in the CMB B-modes and has smaller uncertainties in E-modes.
}
    \end{minipage}

    \vspace{1ex}

    \begin{minipage}{\textwidth}
        \centering
        \includegraphics[width=\textwidth,height=0.28\textheight,keepaspectratio]{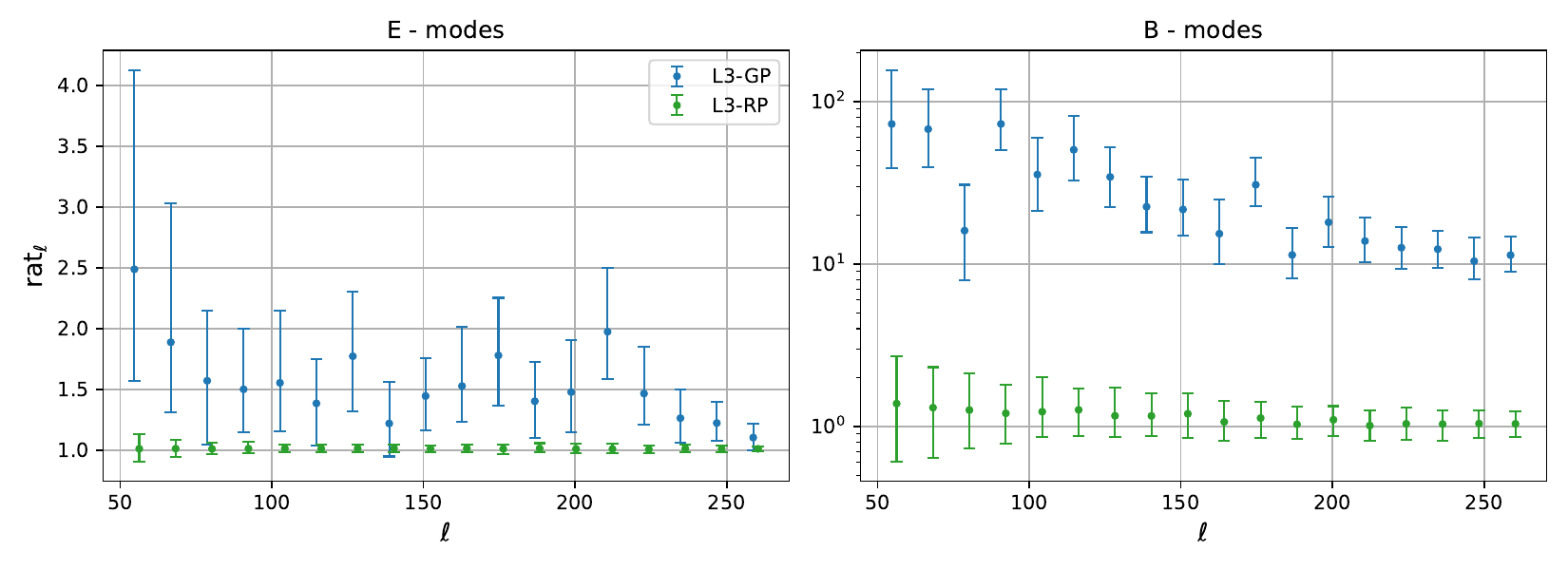}
        \\[-1ex]
        {\small (b) Performance of the CNNs trained on the Rotating Patches (RP) model and the Gaussian Parameters (GP) and tested on the PySM3 model d11s6 model. L3-RP outperforms L3-GP in both CMB E- and B-modes.}
    \end{minipage}

    \vspace{1ex}

    \begin{minipage}{\textwidth}
        \centering
        \includegraphics[width=\textwidth,height=0.28\textheight,keepaspectratio]{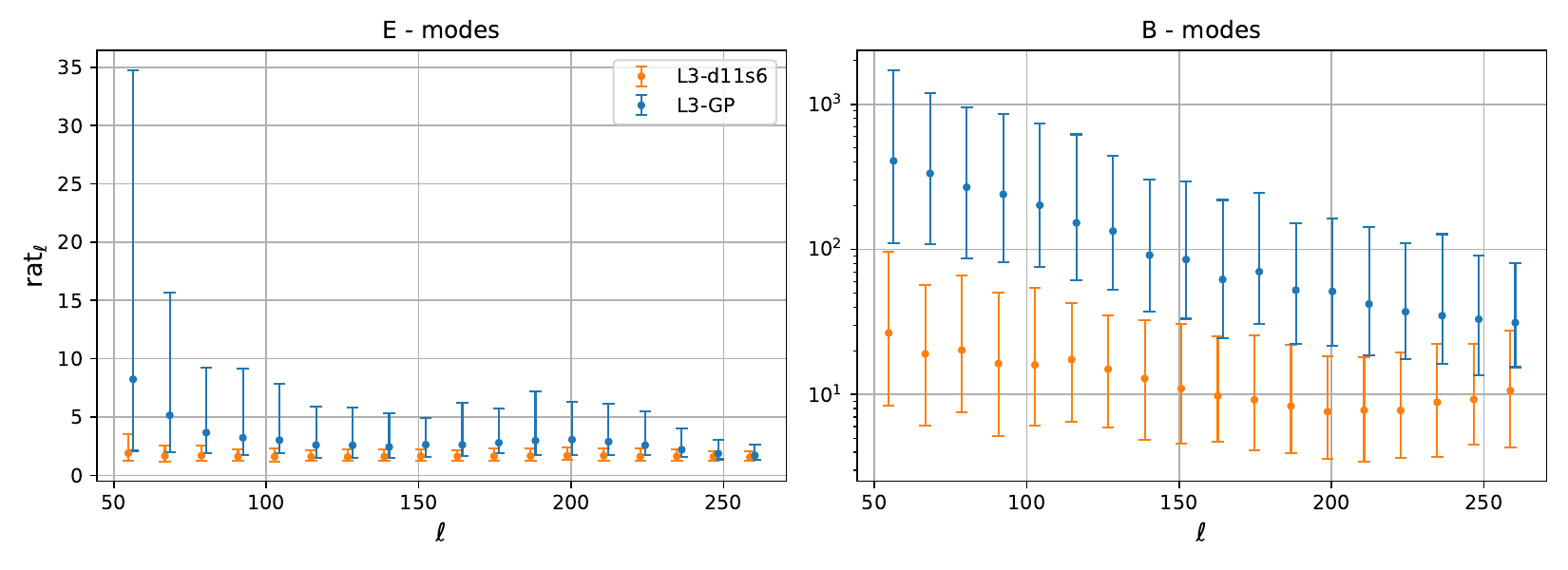}
        \\[-1ex]
        {\small (c) Performance of the CNNs trained on the PySM3 model d11s6 and the Gaussian Parameters (GP) model and tested on the Rotating Patches (RP) model. L3-d11s6 outperforms L3-GP in both CMB E- and B-modes.}
    \end{minipage}

    \caption{Median and 66\% Probability Interval (PI) of the ratio defined in Eq.~\ref{ratio}. 
    Left panels correspond to CMB E-modes, right panels to CMB B-modes. 
    We compare two CNNs tested on a third foreground model.}
    \label{gen_plot}
\end{figure*}

In Fig. \ref{gen_plot}, we show a qualitative comparison of the generalization capabilities of the CNN-based method using the L3 architecture. In each plot, we show the performance of two CNNs tested on the FM that was not used during training.

In Fig.~\ref{gen_plot}.a, we show the results of L3-d11s6 and L3-RP tested on GP. Both for E- and B- modes, L3-RP has smaller error bars and a median closer to one. We therefore conclude that L3-RP performs better on GP than L3-d11s6. In Fig.~\ref{gen_plot}.b, we show the results of L3-GP and L3-RP tested on d11s6. In this case, L3-RP shows better performance, both in terms of the size of the error bars and the median for the E- and B-modes. These two results agree with the prediction of the generalization hypothesis, which implies that training on the FM with higher statistical complexity improves the generalization capabilities of the CNNs. In Fig.~\ref{gen_plot}.c, we show the performance of L3-GP and L3-d11s6 tested on RP. In this case, L3-d11s6 shows better performance than L3-GP for both the E- and B-modes. It is notable that the performance of both CNNs changes dramatically when tested on RP, and that L3-GP also performs poorly when tested on d11s6. This indicates that the design of the training set is a key aspect of the CNN-based method. Moreover, the CNN trained on the most statistically complex model (L3-RP) yields the best results overall, whereas L3-GP yields the worst.

\begin{figure*}[ht!]
    \centering

    \begin{minipage}{\textwidth}
        \centering
        \includegraphics[width=\textwidth,height=0.28\textheight,keepaspectratio]{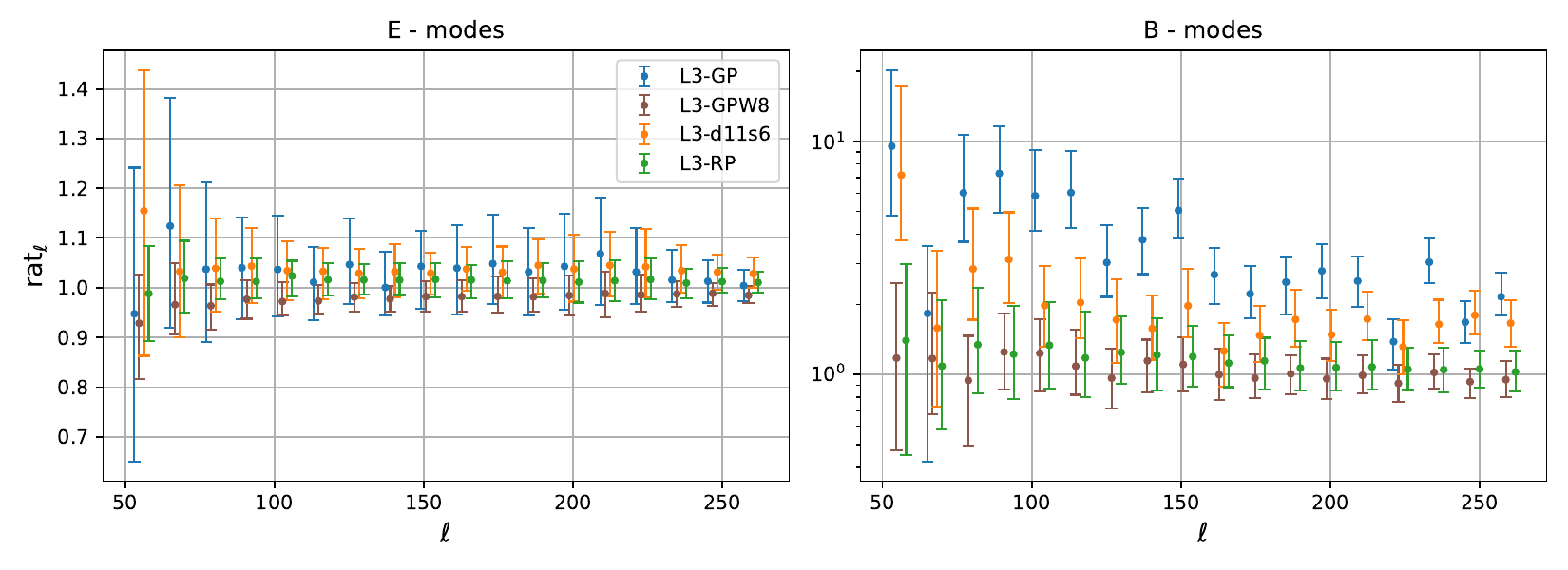}
        \\[-1ex]
        {\small (a) Performance of the CNN-based method tested on the d4s2 model and trained on the four models shown in the legend.}
    \end{minipage}

    \vspace{1.5ex}

    \begin{minipage}{\textwidth}
        \centering
        \includegraphics[width=\textwidth,height=0.28\textheight,keepaspectratio]{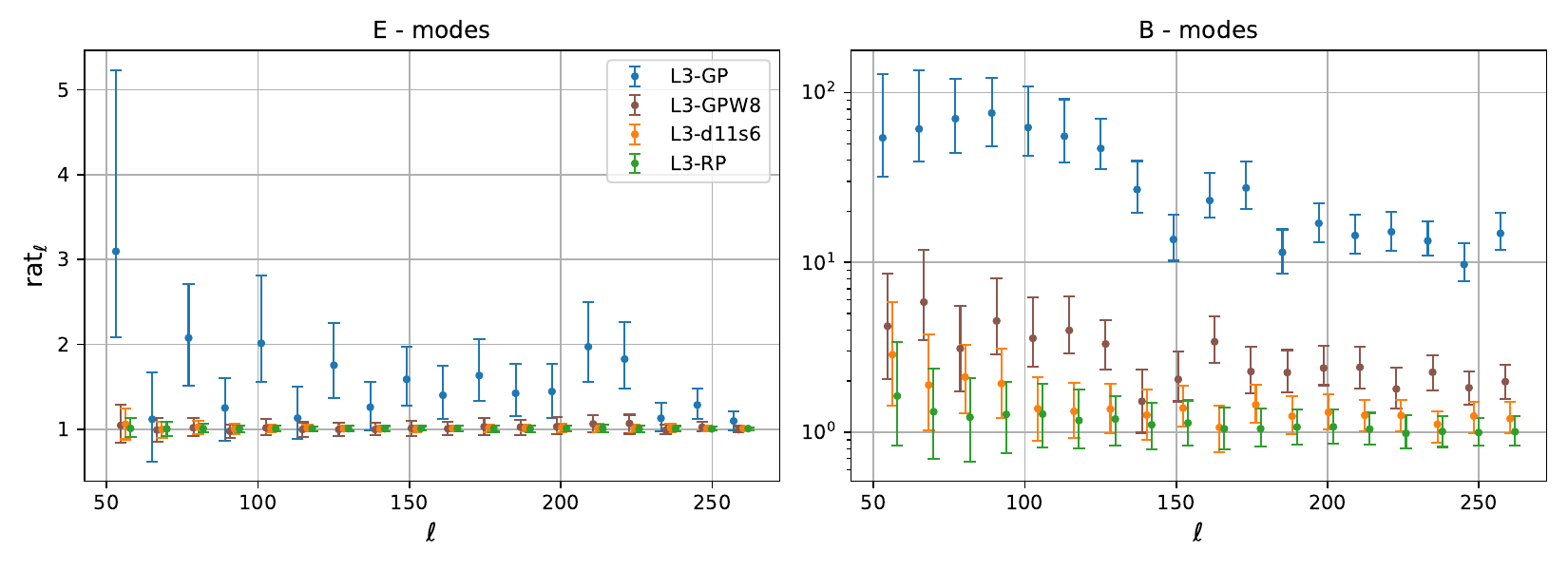}
        \\[-1ex]
        {\small (b) Performance of the CNN-based method tested on the d12s7 model and trained on the four models shown in the legend.}
    \end{minipage}

    \caption{Median and 66\% Probability Interval (PI) of the ratio defined in Eq.~\ref{ratio}. 
    Left panels correspond to E-modes, right panels to B-modes.  
    We compare the four CNNs with two fixed models that were not used during training.}
    \label{comp_over_d4d12}
\end{figure*}

In Fig.~\ref{comp_over_d4d12}, we show the performance of the four CNNs on a fourth and a fifth model, d4s2 and d12s7, respectively, which were not used for the training of the CNNs.

In Fig.~\ref{comp_over_d4d12}.a, we show the results of the four trained L3 architectures on the fixed FM d4s2. In \cite{yan2024cmbfscnn}, it was found that L3-GP shows poor results when tested on d4s2. The authors then add d4s2 simulations to the training set, after which performance improves. In this work, we recover this result and show that L3-RP achieves improved performance without modifying the training set. In this case, we also find that L3-RP outperforms L3-d11s6, as predicted in Section \ref{hierarchy}. In Fig.~\ref{comp_over_d4d12}.b, we show the same CNNs tested on the model d12s7. The performance of L3-GP is the worst among the four, indicating a strong dependence of this CNN on the FM of the testing set. Furthermore, L3-RP performs better than L3-d11s6, supporting our prediction in Section \ref{hierarchy}. L3-GPW8 performs best on d4s2 (averaged $B$-mode bias
$0.09$, against $0.16$ for L3-RP), but not on d12s7, where its bias  ($1.9$)
is an order of magnitude above that of L3-RP ($0.15$).

\subsection{Comparison between the GP modifications}\label{GP_comp_sec}

\begin{figure*}[ht!]
    \centering

    \begin{minipage}{\textwidth}
        \centering
        \includegraphics[width=\textwidth,height=0.28\textheight,keepaspectratio]{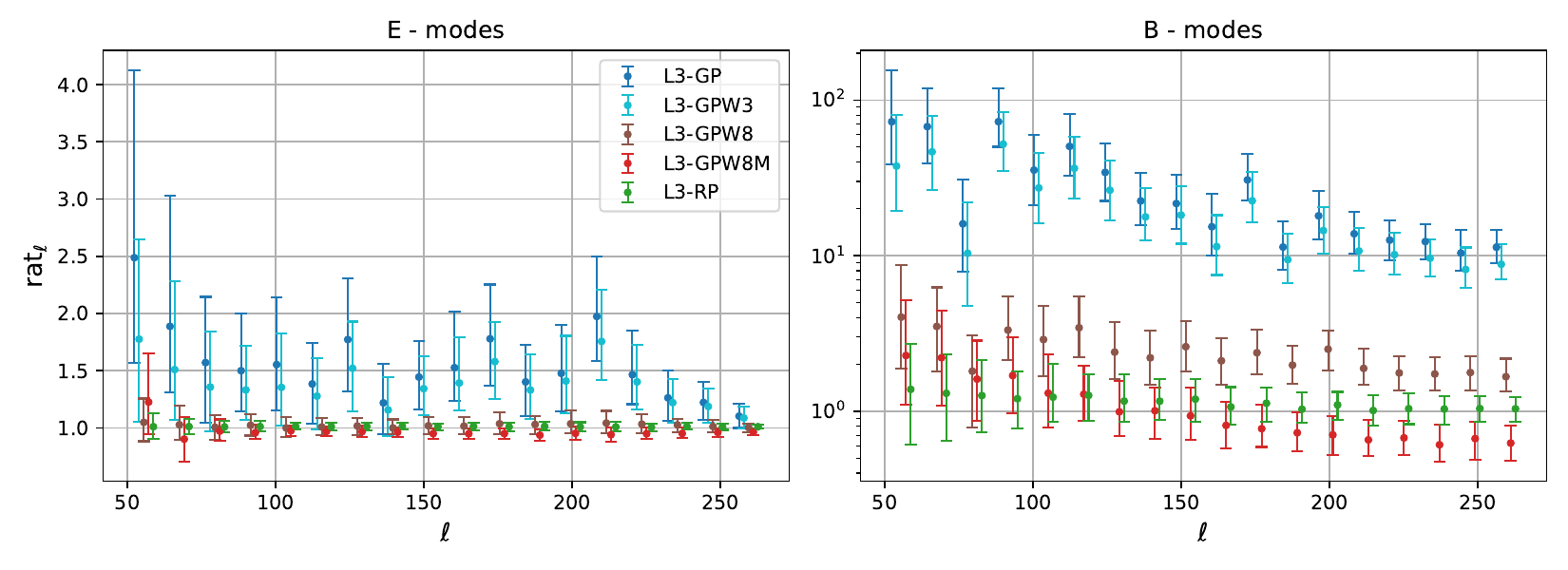}
    \end{minipage}

    \caption{Median and 66\% Probability Interval (PI) of the ratio defined in Eq.~\ref{ratio}. 
    Left panels correspond to E-modes, right panels to B-modes. We compare the modifications of GP and RP, tested on d11s6.}
    \label{wide_d11}
\end{figure*}

Figure~\ref{wide_d11} compares the GP modifications with RP, all of them tested
on d11s6. In this case, increasing the amplitude of the SED variations improves the
reconstruction: the averaged B-mode bias decreases from $28$ for GP to $20$ for GP-W3 and $1.4$ for GP-W8, with a comparable reduction in the precision metric. GP-W8M does not follow this trend: although its marginal statistics are comparable to those of GP-W8, its B-mode bias is lower ($0.40$), but the deviation changes sign across the multipole range instead of decreasing uniformly, and at the highest multipoles its recovered power falls below that of the target, to the level of the CMB alone, so that, unlike the other networks, it does not reproduce the noise power contained in its target; at $\ell<130$, however, its bias is also lower than that of GP-W8 ($0.63$ against $2.1$). In E-modes, GP-W8 and RP are consistent with unity within their uncertainties, so the differences between the models are only visible in B-modes, where none of the GP modifications reaches RP ($0.15$, the smallest bias and scatter among all the networks). Broadening the marginal
distribution and the spectral variability of GP therefore improves generalization, but it does
not reach the performance of RP, which is the most morphologically diverse model by construction.

\subsection{Quantitative generalization analysis}

\begin{figure*}[ht!]
    \centering

    \begin{minipage}{\textwidth}
        \centering
        \includegraphics[width=\textwidth,height=0.25\textheight,keepaspectratio]{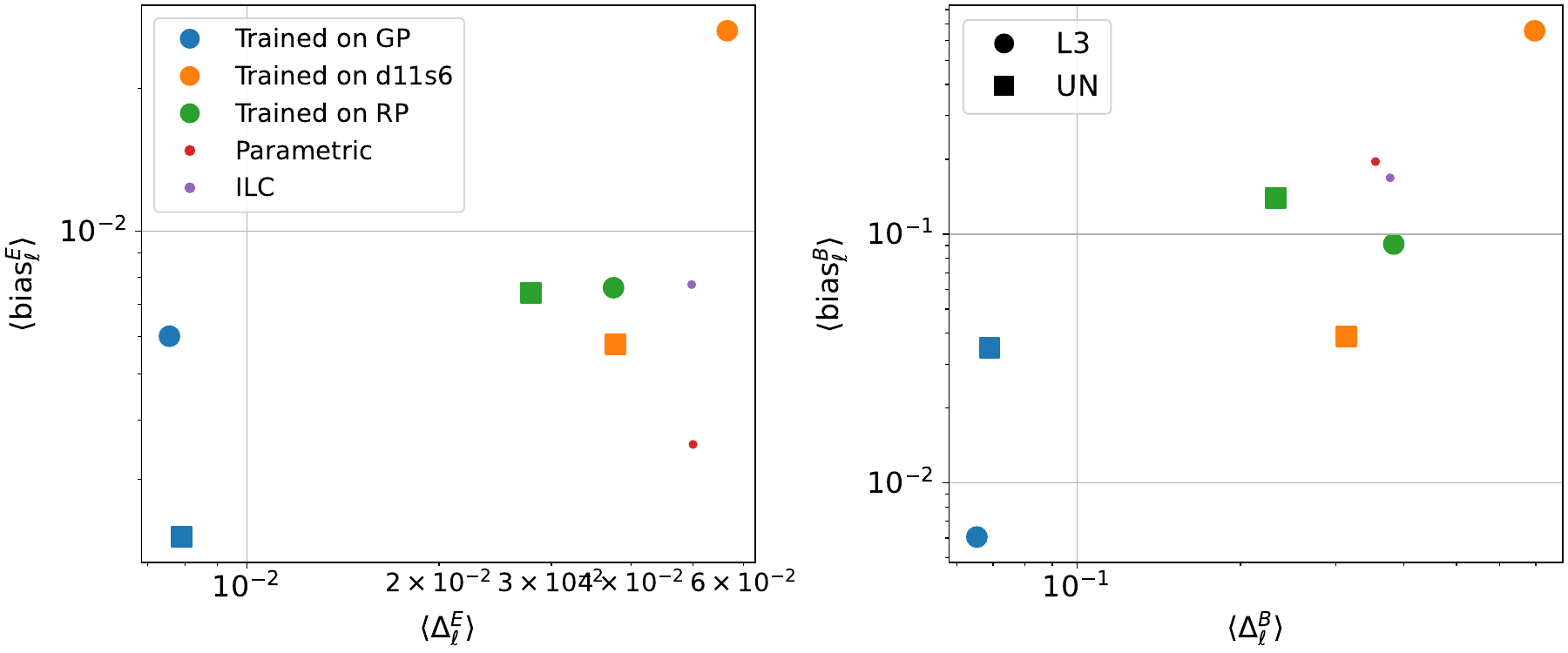}
        \\[-1ex]
        {\small (a) Performance metrics of the methods applied to the Gaussian Parameters (GP) model.}
    \end{minipage}

    \vspace{1.5ex}

    \begin{minipage}{\textwidth}
        \centering
        \includegraphics[width=\textwidth,height=0.25\textheight,keepaspectratio]{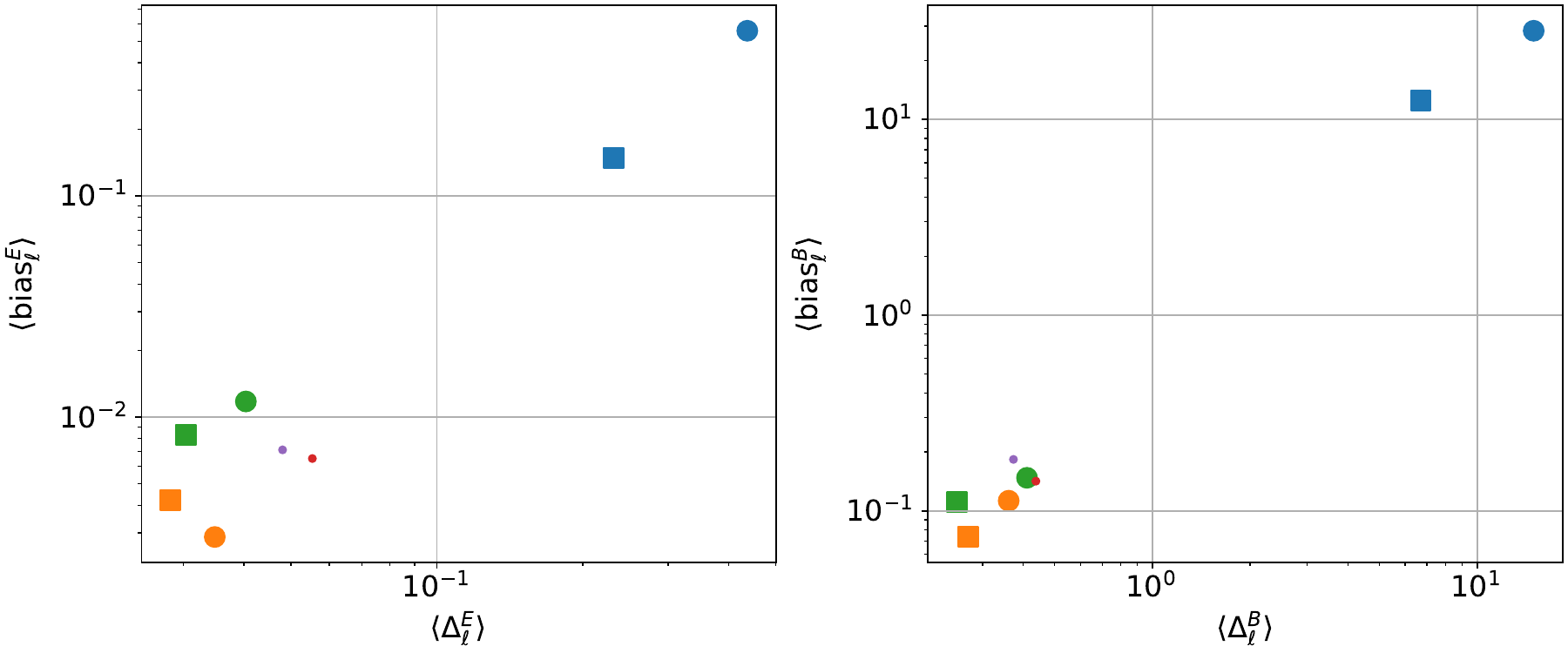}
        \\[-1ex]
        {\small (b) Performance metrics of the methods applied to the PySM3 d11s6 model.}
    \end{minipage}

    \vspace{1.5ex}

    \begin{minipage}{\textwidth}
        \centering
        \includegraphics[width=\textwidth,height=0.25\textheight,keepaspectratio]{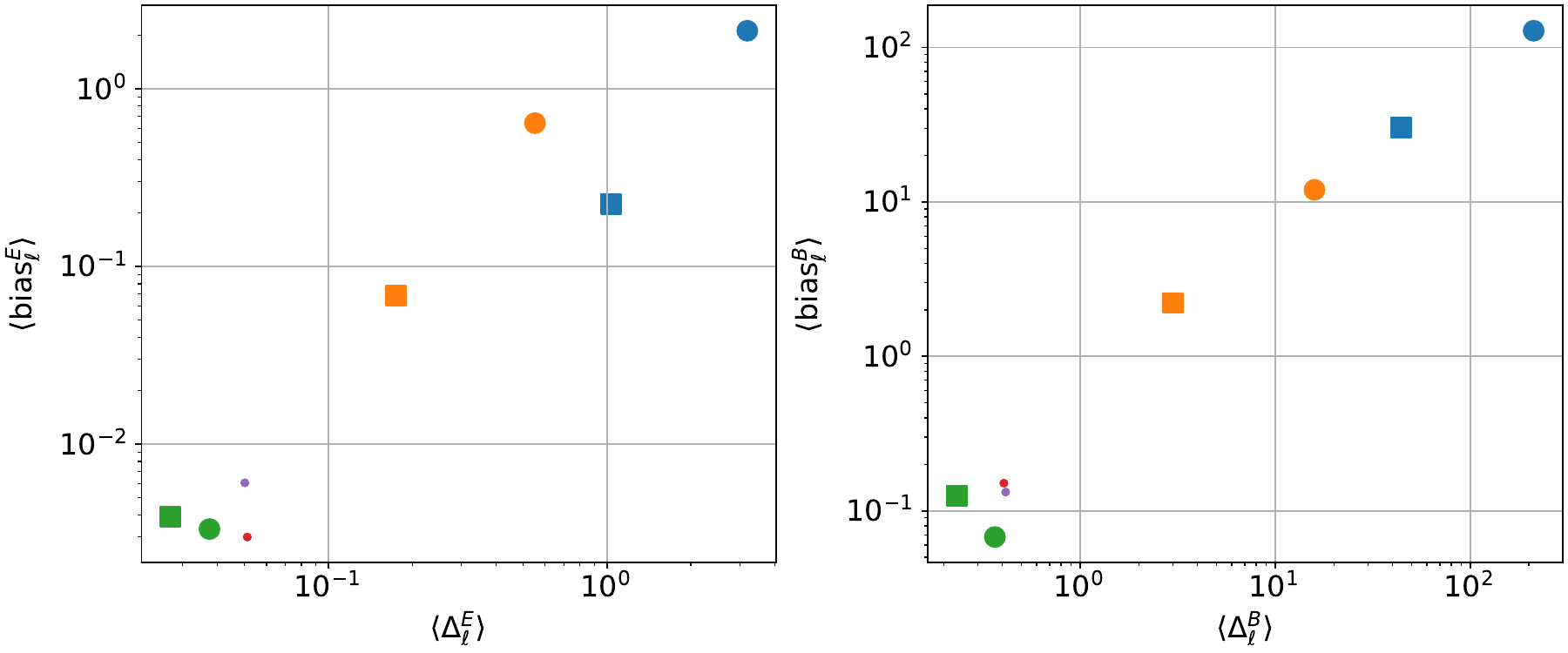}
        \\[-1ex]
        {\small (c) Performance metrics of the methods applied to the Rotating Patches (RP) model.}
    \end{minipage}

    \caption{Accuracy and precision metrics (Eqs. \ref{precision} and  \ref{accuracy}) averaged over the multipole range of interest ($50<\ell<260$). 
    Left panels correspond to CMB E-modes, right panels to CMB B-modes. 
    Each architecture is represented by a different marker, and the foreground model used to generate the training set is represented by a different color.}
    \label{quantitative_analysis}
\end{figure*}

In Fig.~\ref{quantitative_analysis}, we show the averaged precision and accuracy metrics defined in Eqs. \ref{precision} and \ref{accuracy} for a fixed testing set. Fig.~\ref{quantitative_analysis}.a corresponds to the testing set generated with GP, Fig.~\ref{quantitative_analysis}.b to that generated with d11s6, and Fig.~\ref{quantitative_analysis}.c to that generated with RP. We compare different architectures using markers and different colors to identify the FM used to train the CNN. We also include the traditional methods in the plots.

The x-axis of the plots in Fig.~\ref{quantitative_analysis} corresponds to the
averaged precision metric (the size of the error bars), while the y-axis
corresponds to the averaged accuracy metric (the deviation from unity). The
averaged metrics defined in this work allow us to quickly quantify the
generalization capabilities. Both metrics are straightforward to interpret: the
closer to the origin, the better. The results from Section \ref{qualitative_sec}
can be recovered by looking at the circles and comparing green vs. orange (a),
green vs. blue (b), and blue vs. orange (c). By looking at
Figs.~\ref{quantitative_analysis}.a and \ref{quantitative_analysis}.b, we
conclude that the qualitative predictions made for the generalization
capabilities hold for the L3 architecture, and for UN with one exception: on GP, UN-d11s6 has a smaller bias than UN-RP, while UN-RP remains the more precise. The CNNs trained on GP and
d11s6 are not only far from the performance of the CNNs trained on RP: when
tested on a model different from the training one, they can also be worse than the traditional methods in B-modes. This is the case of
L3-GP tested on d11s6 and, more markedly, of both L3-GP and L3-d11s6 tested on
RP, where their bias exceeds that of the ILC by more than an order of magnitude.
Training on a simpler FM can therefore be worse than using a
foreground-model-independent method.

In Table \ref{gen_fact_tab}, we show the results for the generalization factors defined in Eqs. \ref{delta_gen_fact} and \ref{sigma_gen_fact} for the two different architectures used in this work.

\begin{table}[h!]
    \centering
    \begin{tabular}{|c|c|c|c|c|}
    \hline
    &  & [d11s6, RP; GP] & [RP, GP; d11s6] & [GP, d11s6; RP] \\
    \hline\hline
    \multirow{4}{*}{L3}
    & $\mathcal{G}{\rm bias}^E$  & $2\times10^{-2}$ & -0.55 & 1.5 \\
    \cline{2-5}
    & $\mathcal{G}\Delta^E$  & 0.02 & -0.40 & 2.6 \\
    \cline{2-5}
    & $\mathcal{G}{\rm bias}^B$  & 0.57 & -28 & 116 \\
    \cline{2-5}
    & $\mathcal{G}\Delta^B$  & 0.31 & -15 & 196 \\
    \hline
    \hline
    \multirow{4}{*}{UN}
    & $\mathcal{G}{\rm bias}^E$  & $-2\times10^{-3}$ & -0.14 & 0.16 \\
    \cline{2-5}
    & $\mathcal{G}\Delta^E$  & 0.01 & -0.20 & 0.86 \\
    \cline{2-5}
    & $\mathcal{G}{\rm bias}^B$  & -0.10 & -12 & 28 \\
    \cline{2-5}
    & $\mathcal{G}\Delta^B$  & 0.08 & -6.5 & 41 \\
    \hline
\end{tabular}
\caption{Generalization Factor (GF) for the relevant combinations. If a given GF[M$_1$,M$_2$;M$_3$] shows a negative value, it means that the CNN trained on M$_1$ outperforms, and so generalizes better, than the CNN trained on M$_2$, testing on M$_3$. A positive value means the opposite.}
\label{gen_fact_tab}
\end{table}

Table \ref{gen_fact_tab} summarizes the information contained in
Fig.~\ref{quantitative_analysis}. The two architectures agree in the sign of the GFs in all the comparisons except two, both in the bias metrics of the first column, although the magnitudes differ appreciably when testing on RP. By analyzing the sign of the GFs, we conclude that the CNNs trained with RP have better performance than the CNNs trained with d11s6 when tested on GP. The exceptions are $\mathcal{G}{\rm bias}^E_{[{\rm d11s6,RP;GP}]}$ and $\mathcal{G}{\rm bias}^B_{[{\rm d11s6,RP;GP}]}$, where UN-d11s6 has a slightly smaller bias than UN-RP; both UN values ($-2\times10^{-3}$ and $-0.10$) are smaller than the precision metric of these networks, so on GP the two UN networks are comparable in accuracy, while UN-RP remains the most precise. The CNNs trained with RP also outperform the
CNNs trained with GP when tested on d11s6. Finally, the third column shows that
the CNNs trained on d11s6 outperform those trained on GP when tested on RP, with
the largest GF values of the whole table. This is consistent with GP being the
FM with the narrowest training distribution among the three.

\begin{figure*}[ht!]
    \centering
    \includegraphics[width=\textwidth,height=0.28\textheight]{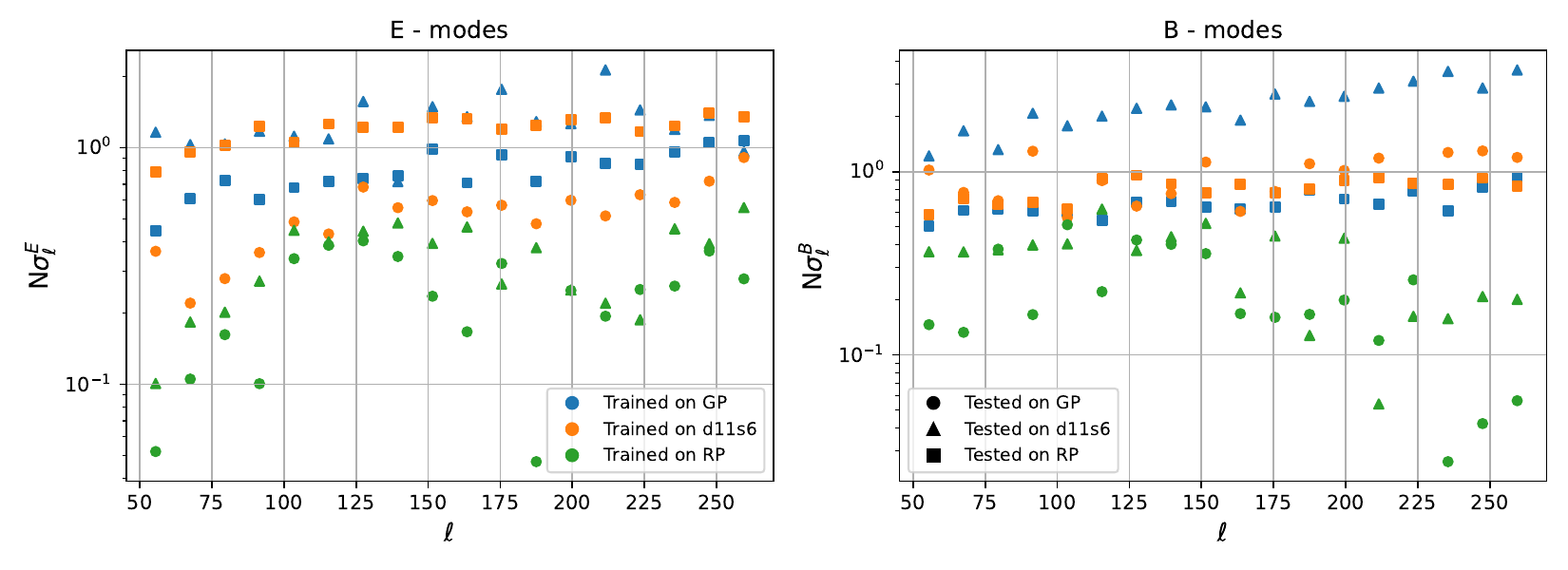}
    \caption{Deviations from a median equal to unity (see Eq.~\ref{nsigm}). Here, we consider the L3 architecture. The color specifies the model assumed during training, while the marker refers to the model used for testing. We do not show CNNs trained and tested on the same foreground model. The y-axis is not a Gaussian sigma, but represents the ratio of median deviation to the 66\% probability interval half-width.}
    \label{nsigmas}
\end{figure*}

In Fig.~\ref{nsigmas}, we show the number of standard deviations that the median
ratio deviates from unity for the L3 architecture when tested on FMs not used
during training. This metric provides a stringent test of the method's
reliability; a value greater than one indicates that the median deviates from unity by more than the half-width of the 66\% probability interval. The largest deviations
occur for L3-GP: in B-modes it exceeds unity over most of the multipole range
when tested on d11s6, reaching N$\sigma\approx3.5$ at the highest multipoles, and in E-modes it reaches N$\sigma\approx2$ around $\ell\approx210$. In contrast, the CNNs trained on RP keep N$\sigma$ well below unity in B-modes over the entire multipole
range, with the smallest values of all the networks around the recombination
bump. In E-modes, L3-GP and L3-d11s6 exceed unity at the highest multipoles when tested on RP (and L3-GP also on d11s6), while the RP-trained networks remain below $\approx0.6$. This is one of the key results of this paper because
the science goal of a SAT, which searches for primordial gravitational waves,
lies in the measurement of the recombination bump, located at $\ell\approx100$.
This is consistent with, and directly anticipates, the bias in the
tensor-to-scalar ratio reported in Section~\ref{t-t-s ratio measure}: the
training model that fails to cover the true B-mode power at the recombination
bump (GP) is precisely the one that yields the largest bias in r
(Fig.~\ref{r_estimation}).

\subsection{Bias in the tensor-to-scalar ratio introduced by the foreground cleaning methods}\label{t-t-s ratio measure} 

\begin{figure}
    \centering
    \includegraphics[width=\linewidth]{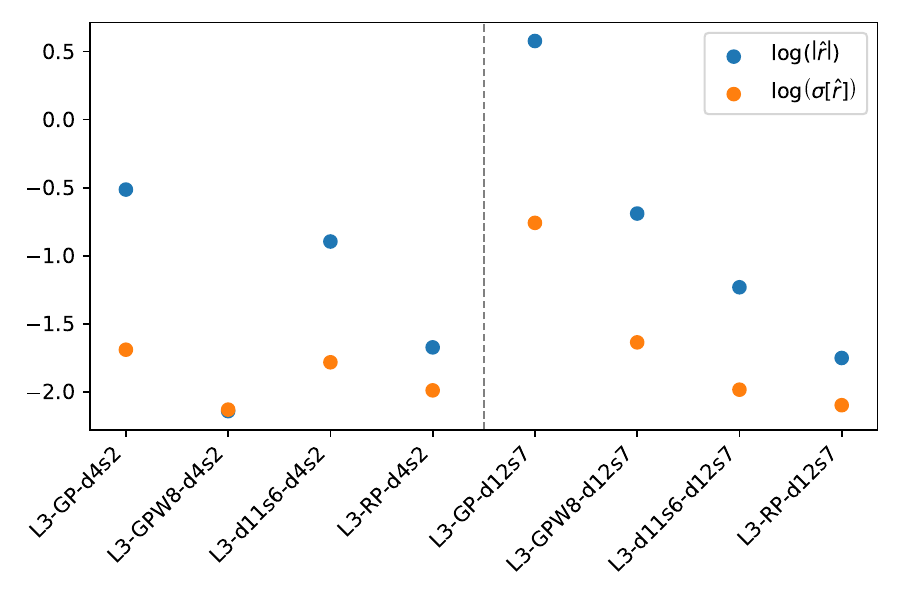}
    \caption{Logarithm of the absolute value of the mean and standard deviation, over the realizations of the testing set,
    of the $r$-equivalent projection of the reconstruction residual $\hat{r}$, for
    the L3 architecture tested on the d4s2 and d12s7 models.}

    \label{r_estimation}
\end{figure}

In Figure \ref{r_estimation}, we show $\log(|\left<\hat{r}\right>|)$ and
$\log(\sigma[\hat{r}])$ for the L3 architecture and the models used only for
testing, d4s2 and d12s7. The estimator $\hat{r}$ is the maximum-likelihood
solution of the model of Eq.~\ref{r_param} under the Gaussian likelihood
described above. Because the lensing term of that model is taken to be the target
spectrum of the same realization, the resulting estimator is the amplitude of the
tensor template that best fits the difference between the reconstructed and the
target spectra, and should therefore be interpreted as an $r$-equivalent measure
of the reconstruction residual rather than as an absolute measurement of $r$;
correspondingly, $\sigma[\hat{r}]$ is the realization-to-realization scatter of
the estimator and not a posterior uncertainty on $r$. No noise debiasing is applied to either spectrum. The target of the network is
the noisy CMB map at $220$~GHz, so $C_\ell^{\rm BB,\,True}$ includes the
instrumental noise power of that channel, and $C_\ell^{\rm BB,\,Rec}$ is the
network's estimate of that same noisy map, obtained from an input that contains
the corresponding noise realization. The noise therefore largely cancels in the
ratio, but not exactly: the residual measured by $\hat{r}$ includes both the
leftover foreground contamination and the error made by the network in
reproducing the noise component, and we do not attempt to separate the two.
 As we noted at the $C_\ell$-level, the CNN trained on RP is the only one whose residual is small on both test models: it shows the smallest residual and scatter on d12s7, and on d4s2 it is exceeded only by L3-GPW8, whose value is discussed below.

The scatter being smaller than the mean value of $\hat{r}$ in all cases but one implies that a systematic residual is introduced by the method. Figure~\ref{r_estimation} shows that training on GP produces a residual equivalent to $r>0.1$ when we remove the foregrounds produced with the model d12s7, which is critical to target primordial gravitational waves. Broadening the training distribution reduces this residual substantially: L3-GPW8 lowers it by a factor of $\approx18$ for d12s7 and $\approx40$ for d4s2, falling below the value obtained with d11s6 on d4s2 but not on d12s7. On d12s7 the smallest residual corresponds to L3-RP, of order $10^{-2}$. On d4s2, L3-GPW8 yields a value ($7\times10^{-3}$) below that of L3-RP ($2\times10^{-2}$) and comparable to its own scatter; its residual changes sign across the multipole range, so the projection onto the tensor template partially cancels, and it should not be read as outperforming L3-RP. Since $\hat{r}$ is a projection of the reconstruction residual and not a full estimate of $r$, these values should not be read as a forecast of the achievable sensitivity; they do indicate, however, that the reconstruction residuals are at a level comparable to current upper limits on $r$.

\section{Conclusion}\label{conc}

In this work, we studied the generalization capabilities of a CNN-based method
for Galactic foreground removal in CMB polarization maps. We show that training
the CNNs assuming the statistically simplest model can introduce systematic
errors in the method, whereas training assuming the statistically most complex
model improves foreground removal performance. The introduced foreground model
(FM) marginal statistics together with the morphological diversity allows us to
define a hierarchy of the breadth of the training
distributions to interpret the
generalization results obtained with the CNN-based method. In particular, the
variance, skewness, and Shannon entropy of the FMs indicate that the Rotating
Patches (RP) foreground model is the most statistically complex model among the
three baseline models considered in this work. We consistently show
that CNNs trained on RP achieve better performance when tested on FMs not used
during training. This result supports our generalization hypothesis: training on
FMs with broader marginal statistics and greater morphological
 diversity leads to improved generalization across different
scenarios, while training on simpler models introduces systematic biases in the
tensor-to-scalar ratio. We further tested this hierarchy directly by
constructing controlled modifications of GP with progressively broader marginal
distributions and spectral variability (GP-W3, GP-W8 and GP-W8M). From GP to GP-W3 and GP-W8, their
statistics move along the same hierarchy, and their generalization performance
improves accordingly, showing that the limitations of GP originate in its
restricted statistical complexity rather than in its parametric construction.
None of these modifications, however, reaches the
robustness of RP: L3-GPW8 outperforms L3-RP on d4s2, but  not on
d11s6 or d12s7. Since GP-W8 exceeds RP in all three marginal
 statistics, marginal variability alone does not determine generalization:
the spatial structure of the training realizations, which the spectral
variability probes only partially, also matters.

Beyond the FM statistics, we introduce quantitative performance metrics to
assess generalization capabilities that are often discussed only qualitatively.
We show that the interpretation of generalization in terms of these metrics is
consistent with the qualitative analysis and provides a new framework for
studying generalization in a fully quantitative manner. These metrics further indicate that the L3 and UN architectures largely agree in the ranking of the training models. We also show the impact on the tensor-to-scalar ratio applied to
FMs not used for the training. This result shows that the L3 architecture trained on RP is the only one that keeps the bias small on both models, while for the d12s7 model, L3-GP reaches a bias larger than $0.1$ for the tensor-to-scalar ratio, which is reduced by a factor of $\approx18$ when the same architecture is trained on GP-W8.

Although CNN-based methods trained on the test model
outperform the traditional foreground-cleaning techniques studied in this work, their results cannot be
considered reliable without a proper understanding of their generalization
properties. In particular, an insufficient understanding of generalization
limits our ability to define robust and meaningful error bars. A clear example
of this limitation is provided by L3-GP, which yields the worst performance when
tested on the PySM3 d11s6, d4s2 and d12s7 models. This behavior highlights the
risks of drawing conclusions from CNN-based methods without a careful assessment
of their range of generalization. Moreover, this result shows that applying a
CNN-based method to real data is not a trivial task, although our
results also indicate a practical mitigation: enlarging the breadth and morphological diversity of the
training FM improves the range of generalization.

\section{Acknowledgements}
\label{sec:Acknowledgements}

This work was supported by the National Science Foundation under Cooperative Agreement PHY-2019786 (The NSF AI Institute for Artificial Intelligence and Fundamental Interactions, http://iaifi.org/). L.G.B. would like to thank Guillermo E. Perna for helpful discussions on the Shannon entropy interpretation.

\bibliographystyle{apsrev4-2}
\bibliography{references}

\appendix
\section{ILC with and without the 30 and 40 GHz frequency bands}

\begin{figure*}[ht!]
    \centering

    \begin{minipage}{\textwidth}
        \centering
        \includegraphics[width=\textwidth,height=0.28\textheight,keepaspectratio]{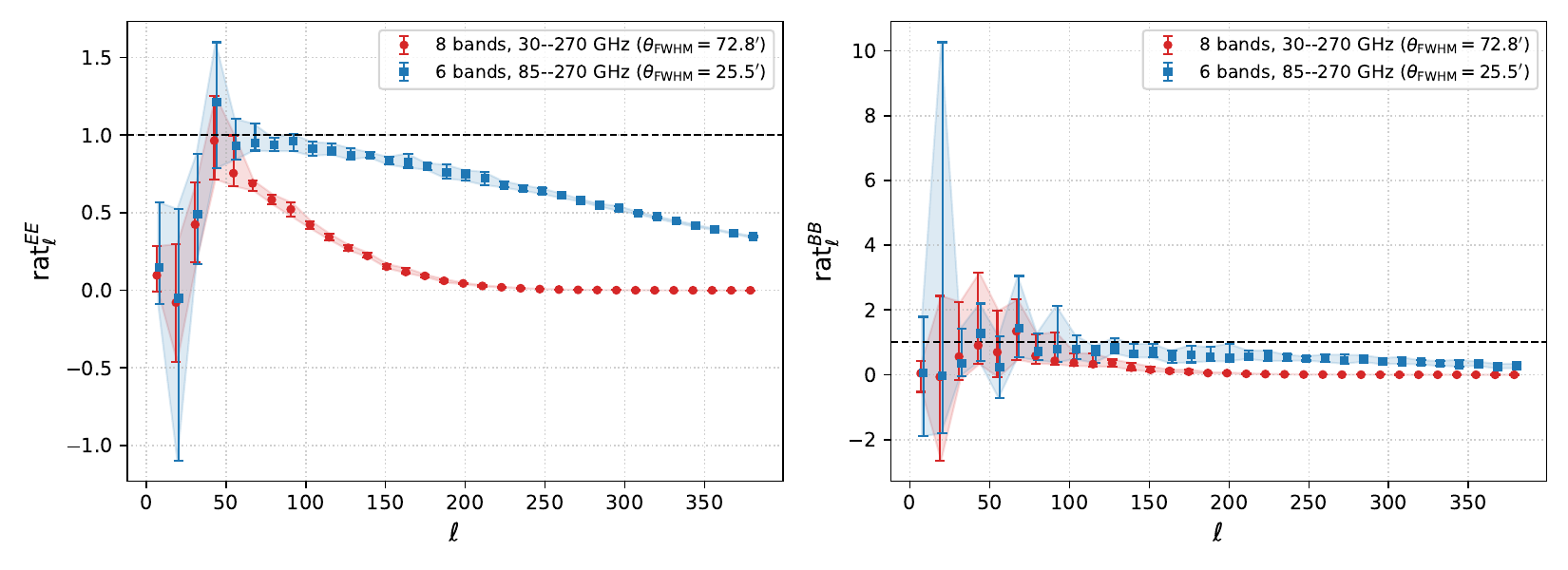}
        \\[-1ex]
    \end{minipage}
     \caption{ILC method with and without the 30 and 40 GHz frequency bands, for 20 realizations of the RP model. Both reconstructions are divided by the same target, the CMB map at the $13'$ beam of the output band, so the decrease with $\ell$ reflects the common beam of each set of inputs ($72.8'$ with the 30 and 40 GHz bands, $25.5'$ without them).}
    \label{apend_ilc}
\end{figure*}

In Figure \ref{apend_ilc}, we show the ILC method applied to our simulations using the RP model for the foregrounds with and without the 30 and 40 GHz frequency bands. 

Convolving the maps to the common frequency beam size of the 30 and 40 GHz implies a loss of small-scale information, which is evident in this result.

\section{Training and validation losses}
\begin{figure*}[ht!]
    \centering
    \begin{minipage}{\textwidth}
        \centering
        \includegraphics[width=\textwidth,height=0.42\textheight,keepaspectratio]{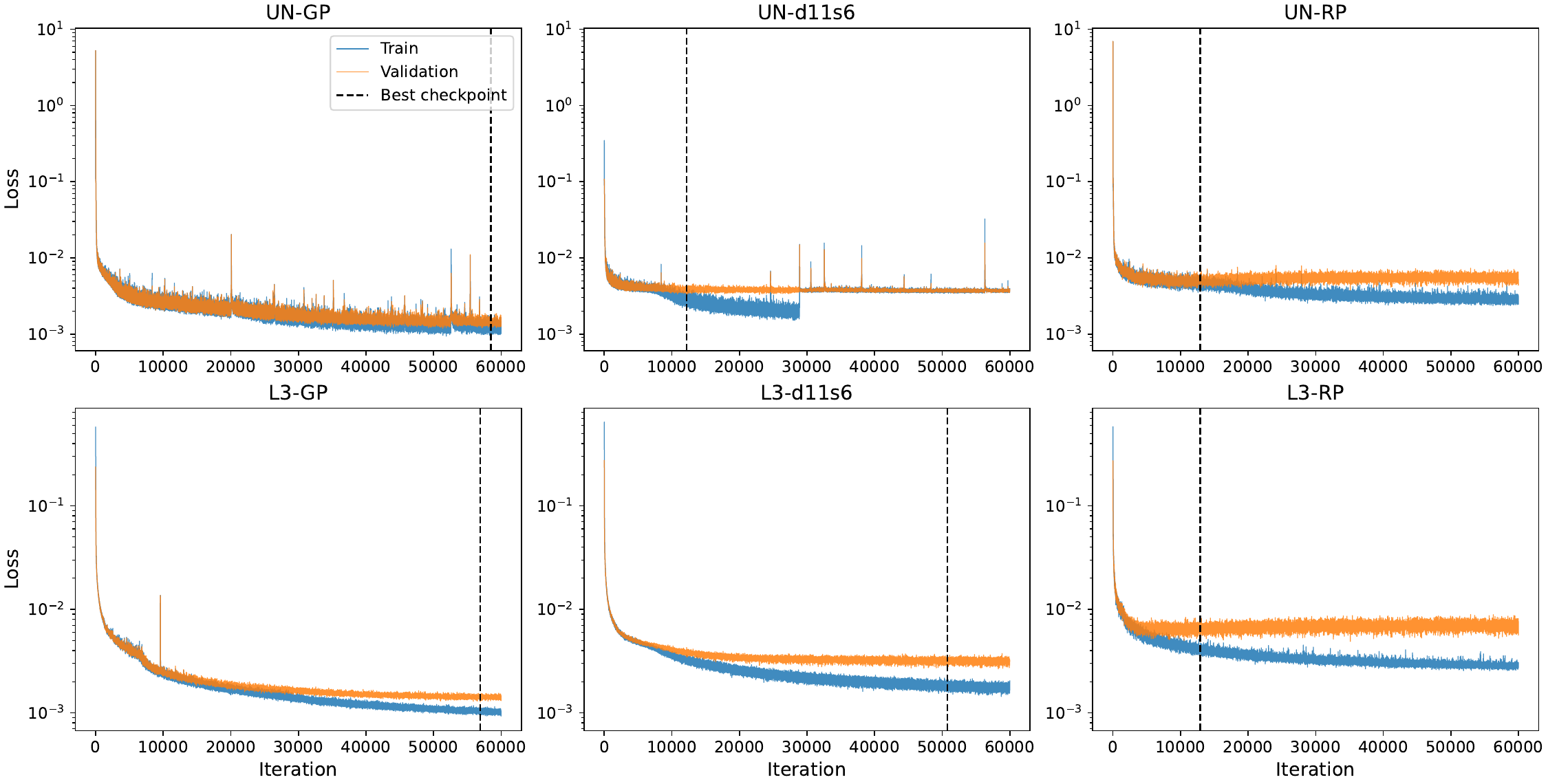}
        \\[-1ex]
    \end{minipage}
    \caption{Training (blue) and validation (orange) loss curves for the UNet
    (top row) and L3 (bottom row) architectures, trained on GP (left), d11s6
    (centre) and RP (right). The dashed vertical line indicates the best
    validation checkpoint, which is the model used throughout this work. The
    two losses separate at late times for all networks except UN-d11s6,
    most markedly for those trained on RP. The UNet trainings show transient
    spikes.}

    \label{apend_loss}
\end{figure*}

In Figure \ref{apend_loss}, we show the training and validation losses for the
six baseline CNNs used in this work: the three UNet models (top row) and the three L3
models (bottom row), each trained on GP, d11s6 and RP. The vertical dashed line
marks the iteration of the best validation checkpoint, which is the set of
weights used to produce every result reported
in this paper. For the UNet, the training and validation losses remain
close for GP and d11s6 at the end of the run (final ratio $1.2$ and
$1.0$, respectively), but separate for RP ($1.9$). The L3 architecture develops a visible
separation between the two curves at late times, with a final ratio, averaged
over the last $5\%$ of the run, of $1.8$ for d11s6, $1.4$ for GP
and $2.5$ for RP. For GP and d11s6 the validation loss
is still decreasing at the end of the run, and the best checkpoint is therefore
reached within the last $15\%$ of the training ($95\%$ and $85\%$
of the total number of iterations, respectively), whereas for RP the validation loss
flattens early and its best checkpoint is
reached at $22\%$. The same holds for UN-RP, whose best checkpoint is also
reached at $22\%$; the RP-trained networks are the only ones whose mean validation
loss then rises, by $7$--$8\%$ between $20$--$30\%$ of the run and its last $15\%$. Since the reported models
are taken at the validation optimum, this late-time separation does not
propagate to any of our results. Retaining the best validation checkpoint
rather than the final weights is a standard early-stopping
procedure~\cite{goodfellow2016}: training is run to completion and
the validation loss is monitored throughout, but the set of weights that is
saved and later used for testing is the one that minimizes it. Every network
used in this work therefore corresponds to its own validation optimum, and none
of them is evaluated in an over-trained regime. The UNet trainings show transient spikes in the loss, after which training
recovers; for UN-d11s6 the
validation optimum ($20\%$ of the run) is attained before the large
transient at $\approx29000$ iterations, and the
subsequent recoveries do not improve upon it.

\end{document}